\documentclass[prb,twocolumn,aps,amsmath,amssymb,floatfix,showpacs]{revtex4-1}

\usepackage[final]{graphicx}
\usepackage[usenames]{color}
\usepackage{afterpage}
\usepackage{epstopdf}
\usepackage{amssymb}
\usepackage{textcomp}
\usepackage{subfigure}
\usepackage{booktabs}
\begin{document}

\title{Projector augmented-wave and all-electron calculations across the periodic
table:\\
a comparison of structural and energetic properties}
\author{
E.~K\"u\c c\"ukbenli$^{1}$}
\thanks{The authors contributed equally}
\author{M.~Monni$^{2,3,*}$}
\author{B.I.~Adetunji$^{4,5}$}
\author{X.~Ge$^{6}$}
\author{G.A.~Adebayo$^{4,5}$}
\author{N.~Marzari$^{1}$}
\author{S.~de~Gironcoli$^{6,7}$}
\author{A.~Dal~Corso$^{6,7}$}
\affiliation{1 Theory and Simulation of Materials, EPFL, CH-1015 Lausanne, Switzerland}
\affiliation{2 Istituto Officina dei Materiali CNR-IOM UOS Cagliari, I-09042 Monserrato, Italy}
\affiliation{3 Dipartimento di Scienze Fisiche, Universit\`a degli Studi di Cagliari, I-09042 Monserrato, Italy}
\affiliation{4 Department of Physics, Federal University of Agriculture, P.M.B. 2240 Abeokuta, Nigeria}
\affiliation{5 The Abdus Salam International Centre for Theoretical Physics, I-34151 Trieste, Italy}
\affiliation{6 International School for Advanced Studies SISSA/ISAS, I-34136 Trieste, Italy}
\affiliation{7 Istituto Officina dei Materiali CNR-IOM DEMOCRITOS, I-34136 Trieste, Italy}

\begin{abstract}
We construct a reference database of materials properties calculated using 
density-functional theory in the local or generalized-gradient approximation, and 
an all-electron or a projector augmented-wave (PAW) formulation, for 
verification and validation of first-principles simulations.
All-electron calculations use the full-potential linearised 
augmented-plane wave method, as implemented in the \texttt{Elk} open-source code,
while PAW calculations use the datasets developed by some of us in the
open-source \texttt{PSlibrary} repository and the \texttt{Quantum ESPRESSO} distribution.
We first calculate lattice parameters, bulk moduli, and energy differences
for alkaline metals, alkaline earths, and $3d$ and $4d$
transition metals in three ideal, reference phases (simple cubic, fcc, and bcc),
representing a standardized crystalline monoatomic solid-state test.
Then, as suggested by K. Lejaeghere {\it et al.}, 
[Critical Reviews in Solid State and Material Sciences
39, p 1 (2014)], we compare the equations of state for all elements,
except lanthanides and actinides, in their experimental phase
(or occasionally a simpler, closely related one).
PAW and all-electron energy differences and structural parameters
agree in most cases within a few meV/atom and a fraction of a percent, respectively.
This level of agreement, comparable with the previous study, includes
also other PAW and all-electron data from the electronic-structure codes
\texttt{VASP} and \texttt{WIEN2K}, and underscores the overall reliability of current, state-of-the-art 
electronic-structure calculations. At the same time, discrepancies that arise even
within the same formulation for simple, fundamental structural properties point to 
the urgent need of establishing standards for verification and validation, reference data sets, and 
careful refinements of the computational approaches used.

\end{abstract}
\pacs{71.15.Nc, 
     71.15.Dx, 
     61.66.Bi 
     }

\maketitle

\section{Introduction}
Density functional theory (DFT) electronic structure calculations within the
plane-wave pseudopotential (PP) method are routinely used to study
and predict the properties of materials as well as for gaining fundamental 
insights in quantum physics and chemistry. Many researchers use 
pseudopotentials to study increasingly complex systems with 
techniques requiring massive numerical efforts such as the many-body 
perturbation theory, time-dependent DFT, crystal structure search,
metadynamics,
high-throughput searches of material properties
\cite{Onida2002,Martonak2006,Curtarolo2013,Jain2013,Marzari2006}.

The transferability of a PP depends on several factors and over the years
many recipes have been proposed, starting from empirical methods 
and arriving to modern ab-initio approaches including 
norm conserving (NC) PPs,~\cite{NC} ultrasoft (US) PPs,~\cite{US} or
the projector augmented-wave (PAW) method.~\cite{Blochl1994,Kresse1999} 
The choice of PP parameters is far from straightforward and in many cases 
requires several refinements through extensive evaluations in which 
PPs are tested in different electronic environments. Moreover, a trade-off between 
accuracy and numerical efficiency must be made leading to PPs of different 
performance. Unfortunately, many pseudopotentials routinely used 
do not come with a standard set of tests and 
must be checked before use, since their accuracy is poorly documented or unknown. 
Moreover, standardized tests that could give an unbiased 
quantitative measure of the PP transferability are still missing, 
with users testing the quantity of interest by searching
reference data in the literature or performing time-consuming all-electron 
calculations for comparison. 
 
Some systematic efforts to obtain a unified, reliable, and accepted test 
procedure for PPs have started to appear in the literature. Standardized set of 
molecules, well established in the 
computational chemistry community (such as the so called G2-1 
set\cite{Curtiss1991,Grossman2002,Nemec2010}) have been used to compare
PAW and all-electron localized basis set 
calculations.~\cite{Paier2005} 
Recently, Lejaeghere {\it et al.}~\cite{Lejaeghere2013} have used
all-electron (\texttt{WIEN2k}\cite{WIEN2kweb}) and
pseudopotential (\texttt{GPAW}\cite{GPAW} and \texttt{VASP}\cite{VASPweb}) codes
to define quantitatively the discrepancies in the equations of 
state fo a wide set of elemental solids in the 
periodic table (71 elements).
In that study, the zero pressure stable phase was chosen as reference for most of the elements. 
More recently, the lattice constant, bulk moduli and energy differences
of the face-centered and body-centered cubic structures of several 
elements have been used to assess the accuracy of a new set of US PPs by 
Garrity, Bennett, Rabe and Vanderbilt (GBVR),~\cite{GBRV} that introduced
a library of PPs targeted at high-throughput calculations.

In the current study, we contribute to these validation and verification efforts
by 
 i) introducing a standardized crystalline monoatomic 
solid test (CMST) where each element is studied in several crystal structures,
ii) providing our own all-electron CMST results using the \texttt{Elk} 
 code\cite{elkweb} and
iii) testing the \texttt{PSlibrary}~\cite{pslibrary} PAW datasets
of the \texttt{Quantum ESPRESSO} (\texttt{QE}) distribution~\cite{espresso} with respect to  \texttt{Elk} for CMST, and 
with respect to \texttt{Elk}, \texttt{WIEN2k}~\cite{WIEN2kweb}, and \texttt{VASP}\cite{VASPweb} 
for the equilibrium structure proposed in
Ref.~\onlinecite{Lejaeghere2013}.

The paper is organized as follows: In Section~\ref{methods} we describe
the methodology of the present study and the computational parameters 
used in our calculations, and we introduce the \texttt{PSlibrary} PAW datasets used
for elements that have not been described elsewhere.
In Section \ref{elk_validation} we validate our \texttt{Elk} all-electron 
results by comparing them, whenever possible to the results in the literature. 
Then, in Section~\ref{paw_validation} \texttt{Elk} data are compared with the 
\texttt{PSlibrary} on the CMST.
In section \ref{paw_test_emine} we follow the methodology of Ref.~\onlinecite{Lejaeghere2013} 
and extend these tests to a majority of the periodic table (68 elements). 
We provide additional \texttt{Elk} and \texttt{QE} results obtained 
using the \texttt{PSlibrary} distribution, which will be referred as \texttt{QE-PAW} hereon, 
to be compared with the previous data.
Conclusions are given in section \ref{conclusions}.

\section{Methodology}
\label{methods}
\subsection{Crystalline Monoatomic Solid Test} 
\label{subsec:cmst}

We propose here to define a standardized {\sl crystalline monoatomic solid test}, CMST, 
consisting of the study of each element in three crystal structures:
simple cubic (\emph{sc}), face-centered cubic (\emph{fcc}), and body-centered cubic (\emph{bcc}), 
focusing on the zero pressure equilibrium lattice constant and bulk modulus and on
the energy differences among the three phases in non-magnetic configurations.

The lattice constant $a_0$, the bulk modulus $B_0$ and the 
total energy $E_{tot}$ at zero pressure are calculated fitting the total 
energy as a function of the volume.
An energy-volume curve is calculated with $15$ points around a first 
estimate of $a_0$ (from -7\% to +7\%, in 1\% steps),
which is fitted with a third order Birch-Murnaghan equation of state (EoS),
\begin{equation}
  \label{eq:E(V)}
  \begin{split}
  E(V) =&E_0 + \frac{9}{16} B_0 V_0 \left[\left(\frac{V_0}{V}\right)^{\frac{2}{3}}-1\right]^2 \\
        &\cdot\left\{2+\left[\left(\frac{V_0}{V}\right)^{\frac{2}{3}}-1\right] \left(B'_0-4\right)\right\},
  \end{split}
\end{equation}
where $E_0$ is the equilibrium total energy , $V_0$ is the equilibrium volume, and 
$B_0$ and $B'_0$ are the bulk modulus and its pressure derivative, respectively.

This procedure is iterated until the new $a_0$ differs from the old 
one by less than $2 \times 10^{-4}$ \AA\  ensuring that the EoS parameters are well converged 
and the initial information on the lattice constant does not bias the final results.
The final EoS fits are then validated by comparing the total energies obtained through direct calculations at $a_0$ and 
the ones obtained from the EoS, which agree very well within 1 meV/atom. 

For a few elements and a few values of the lattice parameter we have encountered convergence issues with 
\texttt{Elk} when the PBE\cite{Perdew1996} exchange-correlation was used, 
while no convergence issues were found in the LDA\cite{Perdew1981} calculations.
In all cases the EoS fit could be carried out satisfactorily as at least 11 points were successfully completed.
The list of elements and structures with convergence issues in PBE are reported in the Supplementary Material.

The CMST procedure was carried out for all elements in the alkaline metals, 
the alkaline earths, and the $3d$ and $4d$ transition metals series.
 
\begin{figure}[ht!]
\includegraphics[width=.99\columnwidth]{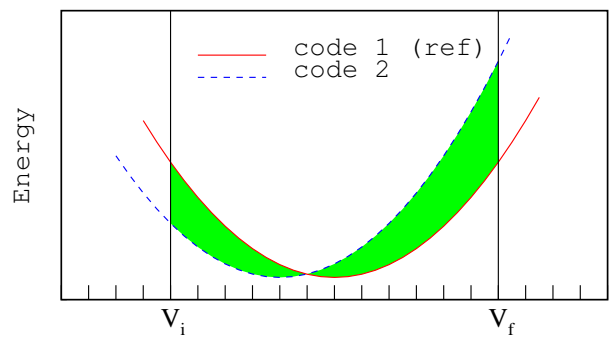}
\caption{Schematic comparison of the EoS obtained by two codes.
Code 1 is taken as a reference and the integration is centered around its 
minimum. 
The $\Delta$ factor is proportional to the mean square deviation of the two 
EoS. The energy difference between the two EoS, whose square enters in the definition of 
$\Delta$, is shown as the shaded area.
}
\label{fig:Delta_eos}
\end{figure}

\subsection{$\Delta$ Factor Calculations}
\label{sec_delta}

In a recent work Lejaeghere {\it et al.} \cite{Lejaeghere2013} propose
to assess the agreement of the DFT description provided by any two 
codes/methods by comparing their resulting equations of state. 
A quality factor $\Delta$, defined by the mean square deviation of the two
EoS, is introduced as
\begin{equation}
\Delta = 
\sqrt{\frac{\int (\Delta E) ^2 (V) dV}{\Delta V}} 
\label{eq:delta}
\end{equation}
where $\Delta E$ is the difference between the energies given by the two
codes/methods and $\Delta V$ is the volume integration interval taken as
$\pm 6$ \% around the equilibrium volume of the code taken as reference.
In order to obtain the EoS they consider seven values around a reference volume $V_{ref}$ 
(0.94 $V_{ref}$ to 1.06 $V_{ref}$) and use the third order Birch-Murnaghan EoS (Eq.~\ref{eq:E(V)}).

In Ref.\onlinecite{Lejaeghere2013}, the comparison is performed among the EoS obtained by 
\texttt{WIEN2k}, \texttt{GPAW} and \texttt{VASP} for a set of 71 elements at their equilibrium structures. 
We extend their work to \texttt{Elk} and \texttt{QE-PAW} 
calculations, using the same procedure, $V_{ref}$ and structure files. 
Furthermore, we also consider a slightly modified definition for the quality factor,
$\Delta'$, that provides a well-defined distance between the codes/methods compared 
(see later Section \ref{paw_test_emine}).

\begin{table*}[t!]
\begin{tabular}{ccccccccc}
\hline
\hline
~&valence&$r_{loc}$ (a.u.) & ~ &$r_c (a.u.) [\epsilon_{ref}$(Ry)$]$ & ~ & $r_{\rm core}$ (a.u.)
&$r_{sph}$ (\AA) & $E_{cut}$ (Ry) \\
~& ~& ~& s  & p & d & ~ & ~ & ~ \\
\hline
Cs & $5s^2 6s^1 5p^6$            &2.2     & 1.9             & 2.3 ~[4.0]      & 2.1 ~[0.3,4.3]  & 1.6  & 1.22 & 35-140 \\
Be & $ 2s^2$                     &1.4 (d) & 2.0 ~[6.5]      & 2.0 ~[6.3]      & 1.4 ~[0.3]      & 1.3  & 1.06 & 30-197 \\
Sc & $3s^2 4s^2 3p^6 3d^1 $      &2.0     & 1.3             & 1.5,~1.8 ~[3.3] & 1.6 ~[0.3]      & 0.8  & 1.06 & 45-477 \\
Ti & $3s^2 4s^23p^64p^03d^2$     &2.1     & 1.7             & 1.8             & 1.7 ~[0.8]      & 0.7  & 1.11 & 50-740 \\
Cr & $3s^2 4s^13p^64p^03d^5$     &2.0     & 1.5             & 1.5,~1.7 ~[3.1] & 1.7 ~[0.3]      & 0.7  & 1.06 & 50-240 \\
Mn & $3s^2 4s^23p^64p^03d^5$     &1.9     & 1.7,~1.6        & 1.9 ~[0.4]      & 1.6 ~[-0.3]     & 0.7  & 1.01 & 46-245 \\
Zn & $3d^{10} 4s^2 4p^0$         &1.7     & 2.2 ~[6.1]      & 2.3 ~[6.3]      & 1.8 ~[-0.6]     & 0.7  & 1.22 & 39-232 \\
Y & $4s^2 5s^24p^65p^04d^1$      &2.4     & 1.6             & 1.7             & 1.9 ~[0.3]      & 1.1  & 1.27 & 36-252 \\ 
Nb & $4s^2 5s^14p^65p^04d^4$     &1.5     & 1.4             & 1.7 ~[-0.5]     & 1.7 ~[0.2]      & 1.0  & 0.90 & 42-365 \\
Mo & $4s^2 5s^14p^65p^04d^5$     &1.4     & 1.3             & 1.7             & 1.7 ~[0.3]      & 1.0  & 0.90 & 50-450 \\
Tc & $4s^2 5s^24p^65p^04d^5$     &1.2     & 1.7,~1.1        & 1.7,~1.2        & 2.0 ~[0.3]      & 0.9  & 1.06 & 62-832 \\
Ru & $4s^2 5s^24p^65p^04d^6$     &1.4     & 1.7,~1.3        & 1.8,~1.3        & 1.8 ~[0.3]      & 0.9  & 0.95 & 62-800 \\
Rh & $4s^2 5s^1 4p^6 5p^0 4d^8$  &1.9     & 1.7,~1.3        & 1.7             & 1.9 ~[0.3]      & 0.9  & 1.01 & 55-439 \\
Cd & $4d^{9.5}5s^25p^{0.5}$      &1.9     & 2.1 ~ [2.3]     & 2.3 ~[8.0]      & 1.8 ~[4.3]      & 1.3  & 1.22 & 33-180 \\
Hf & $5s^2 6s^25p^66p^05d^2$     &2.5     & 1.6             & 1.6             & 1.7 ~[0.3]      & 1.2  & 1.32 & 43-175 \\
Re & $5s^2 6s^25p^66p^05d^5$     &2.1     & 1.4             & 1.5             & 1.55 ~ [0.3]    & 1.1  & 1.11 & 54-216 \\
Os & $6s^2 6p^05d^6 $            &2.4     & 2.4 ~[0.4]      & 2.9 ~[1.6]      & 2.1 ~[0.3]      & 1.31 & 1.53 & 31-124 \\
Hg & $5d^{10} 6s^2 6p^0 $        &1.9     & 2.1 ~[2.3]      & 2.3 ~[8.0]      & 2.2 ~[4.3]      & 1.5  & 1.22 & 29-116 \\
Ge & $3d^{10} 4s^2 4p^2$         &2.2     & 2.1 ~[3.1]      & 2.2 ~[6.3]      & 1.85 ~[-0.4]    & 0.9  & 1.16 & 36-240 \\
As & $4s^2 4p^3$                 &2.0 (d) & 1.8 ~[0.4]      & 2.0 ~[0.4]      & 2.0 ~[0.5]      & 1.6  & 1.06 & 20-103 \\
Se & $4s^2 4p^4$                 &1.9 (d) & 1.8 ~[0.3]      & 2.0 ~[0.3]      & 1.9 ~[0.2]      & 1.4  & 1.06 & 21- 95 \\
Br  & $4s^2 4p^5$                &1.9 (d) & 1.9 ~[-1.0]     & 1.8 ~[0.2]      & 1.9 ~[-0.5]     & 1.4  & 1.01 & 25-108 \\
Sb & $5s^2 5p^3$                 &2.2     & 2.2 ~[5.5]      & 2.3 ~[6.3]      & 2.5 ~[0.1,6.4]  & 1.62 & 1.32 & 33-132 \\
Te & $4d^{10} 5s^2 5p^4$         &2.1     & 2.0 ~[6.3]      & 2.4 ~[6.3]      & 1.7 ~[1.3]      & 0.9  & 1.27 & 43-211 \\
I    & $5s^2 5p^5$               &2.0 (d) & 1.9 ~[-1.0]     & 1.8 ~[0.2]      & 2.0 ~[1.5]      & 1.41 & 1.06 & 24-109 \\
Tl  & $5d^{10} 6s^2 6p^1$        &2.2     & 2.0 ~[5.5]      & 2.6 ~[6.5]      & 2.0 ~[-0.6]     & 1.5  & 1.38 & 35-150 \\
Bi &  $5d^{10} 6s^2 6p^3 $       &2.2     & 2.2 ~[5.4]      & 2.6 ~[6.5]      & 2.0 ~[-1.3]     & 1.4  & 1.38 & 43-172 \\
Po & $5d^{10}6s^2 6p^4 $         &1.8     & 2.0 ~ [1.0]     & 2.2 ~[7.0]      & 1.6 ~[-1.5]     & 1.2  & 1.16 & 53-447 \\
He & $1s^2$                      &1.1 (p) & 1.2 ~ [-0.1]    & 1.1 ~[0.1]      & ~               & ~    & 0.64 & 41-165 \\
Ne & $2s^22p^6$                   &1.1 (d) & 1.5,~1.6 ~[6.3] & 1.5 ~[-0.9]     & 1.1 ~[0.15]     & 0.4  & 0.85 & 37-257 \\
Ar  & $3s^2 3p^6$                &1.8 (d) & 1.8 ~[0.2]      & 2.1 ~[0.2]      & 1.8 ~[0.2,0.45] & 0.9  & 1.11 & 26-225 \\
Kr  & $4s^2 4p^6$                &1.8     & 1.8 ~ [0.3]     & 1.7 ~[0.2]      & 2.0 ~[0.3, 0.7] & 1.4  & 1.06 & 28-220 \\

\hline
\end{tabular}
\caption{Electronic configuration and the matching radii for the local potential ($r_{loc}$),
partial waves ($r_c$), and core density ($r_{\rm core}$), and reference energies ($\epsilon_{ref}$)
used in addition to the energy eigenvalues for the PAW datasets of the \texttt{PSLibrary},
tested in this paper and not reported elsewhere. 
The total radius of the PP (maximum of all matching radii) is given in $r_{sph}$.
The estimates of the wavefunction and density cut-off energies required for these PPs are
reported in the last column. Note that while the density cutoffs are 
rather transferable between different electronic environment calculations, wavefunction cutoffs can 
show a stronger dependence. 
For further, we refer to the \texttt{PSlibrary} files\cite{pslibrary}.
}
\label{pptable}
\end{table*}

\subsection{All-electron FP-LAPW setup}
\label{AE_tech}
\subsubsection{FP-LAPW for CMST}
For the CMST set, we start by calculating the EoS of \emph{fcc}, \emph{bcc} and \emph{sc} phases
varying the muffin-tin radii $R_{MT}$, with a reasonably 
high wavefunction expansion cut off in the interstitial region, 
defined by a maximum K-vector 
length $K_{max}$ (typically such that $R_{MT}K_{max}=9$) and a dense
K-point grid of $12\times12\times12$ with a Gaussian smearing of $0.02$ Ry. 
We then choose the smallest muffin-tin radius that gives a smooth EoS curve. 
With very few exceptions, for a given element we were able to choose the same radius for all 
three phases, which resulted in a higher accuracy in energy differences, thanks to error cancellations. 
In the final calculations the expansion limit, $K_{max}$,
for interstitial wavefunction and the analogous parameter,
${G}_{max}$, for the interstitial density and potential 
were chosen such that the total energy was converged within 1 meV/atom .
The Brillouin zone sampling and Gaussian smearing width were further optimized 
to result in the same convergence in the total energy.
(See Supplementary Material for a list of 
$R_{MT}$, $K_{max}$, ${G}_{max}$, k-grid and smearing width for each element.)

We use the default core-state occupations in \texttt{Elk}. 
For alkaline and alkaline-earth metals we use the default local orbitals provided 
with the exception of Ca, in which adding \emph{d}-like local orbitals 
at an energy slightly above the Fermi energy results in a great improvement 
in the lattice constant prediction.
For transition metals we add \emph{d}-like  and \emph{f}-like local orbitals 
above the Fermi energy, following the suggestion given in Ref.~\onlinecite{Madsen2001}. 

\subsubsection{FP-LAPW for $\Delta$ factor}
In order to calculate the $\Delta$ factor in an unbiased way 
the internal convergence parameters of \texttt{Elk} are set using default values whenever possible. 
In particular, default values for the muffin-tin radius, the core
electron configuration and the local orbitals are used, except for a few elements
(see Supplementary Material for details) where 
convergence and stability issues forced us to optimize $R_{MT}$.

Although the choice of using the default settings may bias results against \texttt{Elk},
we believe that it reflects a realistic situation, where end-users do
not alter the default values unless a convergence or stability issue
arises. The high accuracy limit for \texttt{Elk} has been assessed through CMST instead. 

Basis-set convergence parameters $K_{max}$,
${G}_{max}$, and the Brillouin zone sampling and smearing
width were optimized to ensure convergence of total energies well within 3
 meV/atom. In order to  distinguish metallic and insulating systems
 in an automatic way, an initial scanning of the electronic entropy
with Gaussian smearing was used. 
For resulting metallic systems (except magnetic ones) the Methfessel-Paxton\cite{meth} 
smearing method is employed. 
All calculations are performed with PBE exchange correlation functional 
(See Supplementary Material for a compilation of the converged
parameters).

\subsection{PAW setup}
\label{PAW_tech}
The \texttt{QE-PAW} datasets tested in this paper are distributed 
in the \texttt{QEforge} portal\cite{QEforge} within the
\texttt{PSlibrary} package, version 0.3.1\cite{pslibrary}.
The datasets for Li, Na, K, Rb, Mg, Ca, Sr, Ti, V, Co, Cu, Ga, Ge, Zr,
Nb, Mo, Rh,Pd, Ag, In, Sn, Ba,Ta, W, Ir and Pb
were previously reported in Ref.~\onlinecite{testrel}
while those for H, B, C, N, O, F, Al, Si, P, S, Cl, Fe and Ni 
were described in Ref.~\onlinecite{Waleed}, while Pt and Au were discussed in 
Ref.~\onlinecite{PAWso}.
As a reference we give 
in Table~\ref{pptable} the 
parameters for all the remaining 32 elements used in this work. 
Note that the PP details for Ti, Ge, Nb, Mo, Rh were also given in 
Ref.~\onlinecite{testrel}, but they have been improved by using the
all-electron data calculated in this paper. 
For further details such as augmentation pseudization radii 
we refer to the \texttt{PSlibrary} files\cite{pslibrary}.

\subsubsection{PAW for CMST}
For each element in the CMST, we perform total energy calculations (using LDA or PBE)
 in the \textit{fcc}, \textit{bcc}, and \textit{sc} phases using the procedure discussed in Section \ref{subsec:cmst}.
We then use the EoS fit to determine an estimated equilibrium lattice constant, $a_0$. 
The kinetic energy cut-offs for both the wavefunctions and the charge density are determined by converging the total energy within 0.5 mRy 
at this $a_0$ estimate. 
We then chose 15 points in the $\pm 0.175$ $a.u.$ range around $a_0$, and calculate 
the final energy-volume curve. 
For the Brillouin zone (BZ) integration we use uniform shifted k-point grids of $n\times n\times n$, 
points where $n$ was varied in the range from 4 to 24. 
To increase the convergence rate we use Gaussian or Marzari-Vanderbilt\cite{MV} cold smearing. 
The convergence of the results with respect to the {\bf k}-point sampling
and smearing parameter are tested separately for each system to result in a convergence within 
0.01 \AA~ in the lattice parameter. 
The equilibrium lattice parameter and bulk modulus are calculated from a fit of the energy-volume 
curves to the EoS of Eq.~\ref{eq:E(V)}. 
The quality of the fit is found to be always very high, with a $\chi^2$
lower than $10^{-10}$ Ry$^2$, except in a few cases indicating that the initial estimate for $a_0$
has not been satisfactory. In these cases the procedure is repeated starting
from the new estimate for $a_0$, obtaining finally an accurate
fit.

\subsubsection{PAW for $\Delta$ factor}
For the $\Delta$ factor calculations, each element in the \texttt{PSlibrary} dataset 
is tested in its equilibrium structure as proposed in Ref.~\onlinecite{Lejaeghere2013}.
All calculations are performed with the PBE exchange correlation functional. 

In order to setup the computational parameters for the $\Delta$ factor calculations
the planewave expansion cutoff, the BZ sampling and smearing-width are systematically
varied to result in a 3 meV/atom absolute convergence for the total energy.
As in the case of \texttt{Elk} FP-LAPW calculations an initial scanning of the
electronic entropy with Gaussian smearing is used to identify metallic
systems. For insulators BZ integrations are not critical and the total energy
converges rapidly with respect to the number of k-points, while for metallic systems 
careful smearing\cite{MV}/k-sampling is used to insure convergence
(See Supplementary Material for a compilation of the converged parameters).

\section{Results}

\begin{table*}[ht!]
\centering
\renewcommand{\arraystretch}{1.3}
\begin{tabular}{@{}lrrcrrrcrrr@{}}
\toprule
& \multicolumn{2}{c}{$fcc$} & \phantom{abc}& \multicolumn{3}{c}{$bcc$} & \phantom{abc} & \multicolumn{3}{c}{$sc$}\\
\cmidrule{2-3} \cmidrule{5-7} \cmidrule{9-11}
Element    &  $a_0$ & $B_0$ &&  $a_0$ & $B_0$ & $E_{tot}-E_{tot}^{fcc}$  &&  $a_0$ & $B_0$ & $E_{tot}-E_{tot}^{fcc}$  \\
\midrule
Li & 4.231 [  1] &  15.1 [-0.0] && 3.361 [  2] &  15.2 [-0.0] &  0.001 [  1.1] && 2.665 [  2]  &  13.4 [ 0.2] &  0.120 [ -1.2] \\
Na & 5.111 [ -3] &   9.1 [-0.1] && 4.053 [ -2] &   9.2 [ 0.0] &  0.001 [ -0.7] && 3.293 [ -3]  &   7.3 [-0.1] &  0.124 [ -1.1] \\
K  & 6.358 [  0] &   4.5 [ 0.2] && 5.040 [ -0] &   4.5 [-0.1] &  0.000 [  0.2] && 4.099 [ -2]  &   3.5 [-0.0] &  0.109 [  0.5] \\
Rb & 6.779 [ -7] &   3.6 [ 0.2] && 5.370 [ -2] &   3.6 [ 0.0] &  0.001 [ -4.2] && 4.388 [ -6]  &   2.8 [-0.0] &  0.100 [ -0.2] \\
Cs & 7.344 [-50]&    2.4 [ 0.0] && 5.806 [-35] &   2.4 [ 0.1] &  0.001 [ -3.9] && 4.795 [-39]  &   1.9 [ 0.1] &  0.097 [  4.0] \\
Be & 3.109 [ -8] & 130.7 [ 3.8] && 2.464 [ -8] & 133.0 [ 1.5] &  0.016 [  3.3] && 2.142 [ -5]  &  80.3 [ 0.2] &  0.987 [ 15.4] \\
Mg & 4.435 [ -5] &  38.4 [ 0.8] && 3.512 [ -9] &  38.4 [ 0.8] &  0.016 [ -1.7] && 2.962 [ -5]  &  25.0 [ 0.5] &  0.375 [  7.2] \\
Ca & 5.335 [ -7] &  18.7 [ 0.5] && 4.216 [-10] &  19.0 [ 0.1] &  0.010 [  1.3] && 3.362 [-13]  &  13.3 [-0.6] &  0.379 [  1.5] \\
Sr & 5.794 [ -6] &  14.4 [-0.0] && 4.573 [ -9] &  14.4 [-0.4] & -0.001 [ -0.2] && 3.646 [ -8]  &   8.7 [-0.2] &  0.382 [ -0.0] \\
Ba & 6.032 [-30]&    9.0 [ 0.2] && 4.790 [-22] &  10.1 [ 0.7] & -0.014 [  5.2] && 3.750 [ -9]  &   8.8 [ 0.5] &  0.267 [  9.1] \\
Sc & 4.474 [ -1] &  58.5 [ 0.0] && 3.569 [ -2] &  59.1 [ 0.2] &  0.081 [ -1.5] && 2.851 [ -2]  &  40.2 [-0.1] &  0.701 [ -2.1] \\
Ti & 4.006 [ -2] & 121.9 [ 0.1] && 3.169 [ -2] & 118.3 [ 1.5] &  0.040 [ -1.0] && 2.553 [  2]  &  92.8 [ 1.4] &  0.739 [  6.0] \\
V  & 3.731 [  5] & 199.7 [ 1.7] && 2.932 [ -1] & 206.7 [ 1.1] & -0.295 [  8.9] && 2.386 [  2]  & 164.6 [-2.4] &  0.617 [-52.0] \\
Cr & 3.547 [  4] & 272.0 [ 2.4] && 2.791 [  4] & 295.6 [ 2.8] & -0.414 [  4.2] && 2.280 [  8]  & 218.2 [ 5.0] &  0.595 [ 36.7] \\
Mn & 3.433 [  4] & 327.5 [ 1.9] && 2.726 [  4] & 327.3 [ 1.5] &  0.096 [  4.0] && 2.231 [  3]  & 253.2 [ 2.8] &  0.891 [ -2.0] \\
Fe & 3.378 [  2] & 339.6 [ 3.5] && 2.700 [  2] & 321.5 [ 1.8] &  0.351 [  0.4] && 2.213 [  3]  & 252.1 [ 2.4] &  1.001 [  6.8] \\
Co & 3.378 [  2] & 310.1 [ 7.6] && 2.700 [  2] & 291.9 [ 7.2] &  0.279 [ -4.1] && 2.226 [  3]  & 226.5 [ 5.8] &  0.917 [ 21.1] \\
Ni & 3.424 [  1] & 254.8 [ 6.0] && 2.724 [ -1] & 249.6 [ 6.6] &  0.067 [  0.1] && 2.263 [ -1]  & 187.6 [ 4.7] &  0.741 [ 16.1] \\
Cu & 3.527 [  4] & 185.2 [ 0.9] && 2.803 [  3] & 181.1 [ 2.5] &  0.030 [  8.3] && 2.332 [  3]  & 136.0 [ 2.0] &  0.542 [ -9.6] \\
Zn & 3.796 [-15] &  95.4 [ 4.5] && 3.022 [-10] &  90.5 [ 4.1] &  0.072 [ -5.3] && 2.532 [ -9]  &  65.5 [ 4.9] &  0.242 [ 22.5] \\
Y  & 4.905 [ -0] &  45.2 [ 0.5] && 3.918 [ -5] &  44.3 [-0.6] &  0.117 [  5.5] && 3.135 [ -1]  &  31.4 [-0.5] &  0.759 [ -0.4] \\
Zr & 4.422 [  5] & 101.5 [ 1.1] && 3.486 [  2] &  96.8 [ 2.8] &  0.019 [ -0.8] && 2.835 [  8]  &  80.4 [ 0.1] &  0.810 [  1.3] \\
Nb & 4.138 [  5] & 177.7 [ 3.9] && 3.250 [  0] & 194.4 [-7.7] & -0.330 [-33.8] && 2.665 [  2]  & 147.7 [ 0.8] &  0.626 [ 24.0] \\
Mo & 3.942 [ -2] & 263.9 [ 5.6] && 3.115 [ -1] & 287.1 [ 7.6] & -0.438 [  5.7] && 2.556 [  2]  & 217.7 [ 2.0] &  0.744 [ 12.1] \\
Tc & 3.816 [ -2] & 333.5 [ 7.7] && 3.036 [ -2] & 329.5 [ 7.7] &  0.201 [ -1.3] && 2.491 [ -2]  & 254.2 [ 3.2] &  1.010 [ 19.8] \\
Ru & 3.755 [ -5] & 341.3 [19.3] && 3.006 [ -3] & 309.5 [22.3] &  0.557 [  5.1] && 2.468 [ -1]  & 258.2 [ 5.2] &  1.102 [  6.2] \\
Rh & 3.758 [  5] & 312.9 [ 4.8] && 3.010 [  4] & 285.8 [ 4.2] &  0.412 [ -6.6] && 2.482 [  4]  & 227.2 [ 3.0] &  0.886 [ 16.2] \\
Pd & 3.846 [  2] & 223.6 [ 7.5] && 3.060 [  2] & 216.7 [ 7.5] &  0.063 [  6.4] && 2.542 [  2]  & 163.3 [ 7.5] &  0.606 [  9.8] \\
Ag & 4.011 [  2] & 136.1 [ 3.5] && 3.191 [ -1] & 132.9 [ 6.8] &  0.035 [ -8.4] && 2.653 [  4]  & 103.0 [ 1.0] &  0.411 [ 18.0] \\
Cd & 4.321 [-14] &  66.8 [ 2.8] && 3.446 [ -9] &  59.6 [ 4.4] &  0.067 [  0.9] && 2.862 [ -6]  &  50.4 [ 2.0] &  0.169 [  2.2] \\
\bottomrule
\end{tabular}
\caption{LDA results for the equilibrium lattice constant $a_0$, bulk modulus $B_0$ and total energy difference
with the \emph{fcc} phase (expressed in \AA, GPa and eV respectively) calculated with LAPW for the solids included in CMST. 
Deviation from LAPW of the PAW results are reported in square brackets (expressed in m\AA, GPa and meV respectively). 
}
 \label{a0_B0_LDA}
\end{table*}

\begin{table*}[ht!]
\centering
\renewcommand{\arraystretch}{1.3}
\begin{tabular}{@{}lrrcrrrcrrr@{}}
\toprule
& \multicolumn{2}{c}{$fcc$} & \phantom{abc}& \multicolumn{3}{c}{$bcc$} & \phantom{abc} & \multicolumn{3}{c}{$sc$}\\
\cmidrule{2-3} \cmidrule{5-7} \cmidrule{9-11}
Element    &  $a_0$ & $B_0$ &&  $a_0$ & $B_0$ & $E_{tot}-E_{tot}^{fcc}$  &&  $a_0$ & $B_0$ & $E_{tot}-E_{tot}^{fcc}$  \\
\midrule
Li &  4.324 [ 1] & 13.9 [-0.1] &&  3.433 [ 3] & 14.0 [-0.0] &  0.001 [ 1.2] &&  2.733 [-0]  & 12.5 [-0.3] &  0.119 [ 2.2] \\
Na &  5.298 [-5] &  7.7 [-0.0] &&  4.201 [-3] &  7.7 [-0.0] & -0.001 [ 1.7] &&  3.416 [-6]  &  6.1 [ 0.1] &  0.120 [-1.4] \\
K  &  6.666 [-2] &  3.5 [ 0.1] &&  5.284 [ 0] &  3.6 [ 0.0] & -0.001 [ 1.7] &&  4.299 [-1]  &  2.8 [ 0.0] &  0.104 [ 0.6] \\
Rb &  7.160 [-12] &  2.8 [ 0.1] &&  5.674 [-0] &  2.8 [-0.1] & -0.001 [-3.1] &&  4.630 [-4]  &  2.1 [-0.0] &  0.095 [ 2.5] \\
Cs &  7.815 [-30] &  1.9 [-0.0] &&  6.193 [-19] &  1.9 [ 0.0] &  0.001 [-2.7] &&  5.073 [-21]  &  1.5 [-0.0] &  0.089 [ 3.5] \\
Be &  3.157 [-5] & 119.8 [ 0.3] &&  2.501 [-5] & 124.1 [ 0.7] &  0.017 [ 1.5] &&  2.174 [-4]  & 75.6 [-0.4] &  0.923 [ 9.6] \\
Mg &  4.519 [-1] & 35.7 [-0.4] &&  3.577 [-4] & 35.6 [-0.4] &  0.018 [-3.6] &&  3.018 [ 1]  & 22.9 [-0.3] &  0.369 [ 1.8] \\
Ca &  5.533 [-16] & 17.1 [ 0.3] &&  4.385 [-5] & 16.2 [ 0.3] &  0.016 [ 0.1] &&  3.521 [-11]  & 11.2 [-0.4] &  0.391 [ 3.6] \\
Sr &  6.029 [-5] & 11.6 [-0.2] &&  4.765 [-8] & 11.7 [-0.0] &  0.006 [ 0.1] &&  3.852 [-9]  &  7.3 [-0.2] &  0.383 [ 4.2] \\
Ba &  6.399 [-37] &  8.0 [ 0.1] &&  5.053 [-25] &  8.5 [ 0.4] & -0.017 [-0.1] &&  3.981 [-14]  &  7.3 [ 0.9] &  0.285 [ 6.3] \\
Sc &  4.621 [-1] & 51.0 [ 0.0] &&  3.678 [-1] & 53.0 [ 0.4] &  0.054 [ 1.1] &&  2.961 [ 5]  & 34.5 [ 0.2] &  0.720 [-1.4] \\
Ti & 4.114 [ -1] & 107.0 [ 0.9] &&  3.256 [-1] & 105.4 [ 0.6] &  0.057 [-4.0] &&  2.638 [ 4]  & 76.4 [ 2.5] &  0.778 [ 3.0] \\
V  &  3.820 [ 1] & 174.0 [ 1.7] &&  2.997 [ 3] & 181.7 [-0.8] & -0.244 [ 1.7] &&  2.447 [ 4]  & 137.4 [ 0.8] &  0.596 [ 2.8] \\
Cr &  3.624 [ 3] & 236.0 [ 0.9] &&  2.851 [ 1] & 256.2 [ 1.6] & -0.393 [ 2.5] &&  2.340 [ 3]  & 186.9 [ 0.2] &  0.634 [-0.2] \\
Mn &  3.506 [ 2] & 278.3 [ 1.8] &&  2.785 [ 3] & 276.6 [ 0.9] &  0.081 [ 2.0] &&  2.285 [ 1]  & 210.1 [ 2.9] &  0.866 [ 1.0] \\
Fe &  3.452 [ 2] & 283.8 [ 1.9] &&  2.761 [ 1] & 266.3 [ 2.1] &  0.312 [ 4.5] &&  2.269 [ 1]  & 206.2 [ 1.5] &  0.948 [ 7.6] \\
Co &  3.458 [ 3] & 252.9 [ 4.8] &&  2.765 [ 2] & 237.4 [ 4.5] &  0.259 [-10.6] &&  2.284 [ 3]  & 182.2 [ 3.6] &  0.856 [15.5] \\
Ni &  3.516 [-1] & 201.7 [ 3.4] &&  2.797 [-2] & 197.8 [ 3.5] &  0.055 [ 3.7] &&  2.327 [ 0]  & 146.3 [ 2.5] &  0.661 [21.5] \\
Cu &  3.635 [11] & 139.8 [-1.1] &&  2.890 [ 8] & 136.7 [-2.9] &  0.033 [ 1.5] &&  2.410 [ 8]  & 101.5 [-1.4] &  0.465 [-3.6] \\
Zn &  3.940 [-14] & 67.2 [ 1.5] &&  3.139 [-8] & 61.4 [ 3.2] &  0.062 [-0.6] &&  2.639 [-8]  & 45.7 [-2.3] &  0.192 [62.0] \\
Y  &  5.067 [-3] & 39.3 [ 0.1] &&  4.043 [-0] & 38.9 [-0.1] &  0.099 [-2.5] &&  3.264 [ 4]  & 25.9 [-0.0] &  0.772 [-0.1] \\
Zr &  4.531 [ 2] & 89.7 [ 0.8] &&  3.577 [ 2] & 86.7 [ 0.6] &  0.054 [-10.6] &&  2.916 [ 3]  & 68.6 [ 1.9] &  0.834 [ 2.5] \\
Nb &  4.223 [-4] & 160.2 [ 4.1] &&  3.314 [-3] & 166.3 [ 7.5] & -0.321 [-2.7] &&  2.722 [-2]  & 127.7 [ 3.2] &  0.649 [ 7.9] \\
Mo &  4.007 [-4] & 234.1 [ 5.3] &&  3.165 [-3] & 254.6 [ 5.4] & -0.424 [-3.7] &&  2.609 [-8]  & 184.9 [14.1] &  0.723 [-2.0] \\
Tc &  3.876 [-3] & 291.7 [ 8.1] &&  3.084 [-3] & 287.7 [ 5.8] &  0.182 [-0.4] &&  2.535 [-2]  & 217.5 [ 4.1] &  0.965 [12.2] \\
Ru &  3.814 [-2] & 302.5 [ 3.5] &&  3.055 [-0] & 276.8 [ 5.2] &  0.512 [ 3.2] &&  2.513 [ 0]  & 217.5 [ 3.0] &  1.015 [ 5.6] \\
Rh &  3.833 [13] & 253.7 [-0.1] &&  3.073 [10] & 230.1 [ 3.5] &  0.362 [-11.8] &&  2.538 [10]  & 180.4 [ 2.2] &  0.789 [-10.1] \\
Pd &  3.946 [ 4] & 166.5 [ 1.2] &&  3.141 [ 3] & 162.0 [ 1.0] &  0.042 [-2.0] &&  2.617 [ 3]  & 119.4 [ 1.8] &  0.499 [-1.6] \\
Ag &  4.155 [ 1] & 88.6 [ 6.4] &&  3.306 [-2] & 87.8 [ 3.8] &  0.025 [-4.4] &&  2.757 [-2]  & 65.0 [ 3.3] &  0.319 [27.0] \\
Cd &  4.519 [-17] & 39.8 [ 2.4] &&  3.614 [-13] & 34.3 [ 2.9] &  0.048 [ 1.9] &&  3.007 [-11]  & 28.2 [ 2.3] &  0.108 [10.9] \\
\bottomrule
\end{tabular}
\caption{PBE results for the equilibrium lattice constant $a_0$, bulk modulus $B_0$ and total energy difference
with the \emph{fcc} phase (expressed in \AA, GPa and eV respectively) calculated with LAPW for the solids included in CMST. 
Deviation from LAPW of the PAW results are reported in square brackets (expressed in m\AA, GPa and meV respectively). 
}
\label{a0_B0_PBE}
\end{table*}

\subsection{Comparison of LAPW data with literature results}
\label{elk_validation}
Motivated by the technical nature of the calculations, 
before providing our findings, we first verify 
our LAPW calculations with existing results in the literature,
given that some of the crystals included in the CMST were present in published work
\cite{Haas2009,Haas2009err,Tran2007}. For example Ref.~\onlinecite{Haas2009}
shares with our test 16 cases out of 90. The relative lattice constant difference between these
is less than 0.16\%; exceptions are Fe $bcc$ which shows a large discrepancy
(around 2\% both for LDA and PBE) and Ba $bcc$ (-0.76\% in LDA). Ref.~\onlinecite{Tran2007}
reports the bulk modulus of the same systems: the lattice constants can be seen to be identical to Ref.~\onlinecite{Haas2009}
(apart for those specified in the errata\cite{Haas2009err}).
The  relative difference in bulk moduli between Ref.~\onlinecite{Tran2007} and 
our present results is less than 3.2\%, the only exception being Fe $bcc$.
These differences between two independent FP-LAPW calculations give an estimate of a lower bound 
for the accuracy to be targeted in a PAW and FP-LAPW comparison. 

Unfortunately other all-electron data available in literature are limited to a small number 
of crystalline systems and/or are calculated with different techniques.
As an example, Ref.~\onlinecite{Ropo2008} reports exact muffin-tin orbitals (EMTO) results. EMTO is
 a method of the Korringa-Kohn-Rostoker type, credited to be as accurate as full-potential methods.
\cite{Ropo2008,Asato1999}
Ref.~\onlinecite{Ropo2008} and this study have 18 systems in common. The relative lattice constant
difference is less than 0.5\%; exceptions being Fe, Ba and Cs
$bcc$ in LDA,  and Ni $fcc$ in PBE.
The relative difference in bulk moduli is less than 5\% in approximately half of the cases and 
larger in the remaining cases, up to 28\% for Fe $bcc$ in PBE. 
These relative differences are greater than the ones found with respect
to the FP-LAPW calculations,\cite{Haas2009,Tran2007} and are of the same order as the difference between other 
LAPW results\cite{Haas2009,Tran2007} and the EMTO results.  \cite{Ropo2008}

Another example is Ref.~\onlinecite{Mattsson2008}, which reports calculations performed with the
full-potential linear muffin-tin orbital method (LMTO).
Only 6 systems are common and the results are quite close 
both in terms of $a_0$ and $B_0$, differences being 
at most 0.3~\% in the lattice parameter and 2~\% in the bulk modulus.

The \texttt{Elk} FP-LAPW CMST database expands and fills the  
gaps in all-electron data available in the literature providing an extensive set of
systems, generated by a uniform procedure in terms of computational method
used and convergence parameters, that includes several
simple structures for each element and reports both structural properties and energy
differences. 
While the choice of cubic structures for CMST is at variance with present tests in the literature that
mostly focus on the experimental stable phases, 
it allows an extensive comparison of the energetics
of $fcc$, $bcc$ and $sc$ phases which we believe
are of interest for rapidly assessing the performance of a computational methodology
in the most homogeneous way. 

\subsection{CMST comparison between PAW and FP-LAPW data}
\label{paw_validation}
\begin{figure}[ht!]
\includegraphics[width=.99\columnwidth]{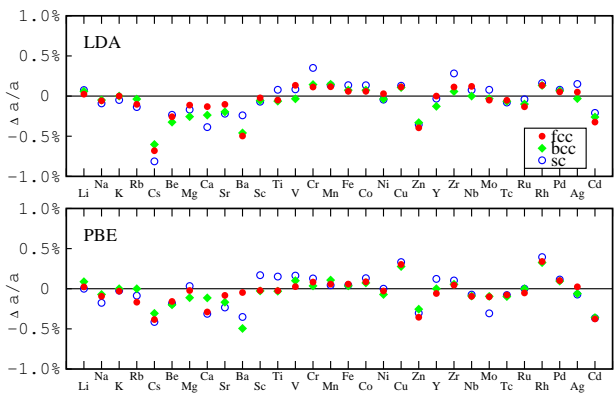}
\caption{Percentage differences between LAPW and PAW lattice constants  
for the solids included in CMST, in all three crystalline phases considered,
calculated using LDA (top ) and PBE (bottom) functionals. 
}
\label{fig_a0}
\end{figure}

\begin{figure}
\includegraphics[width=0.99\columnwidth]{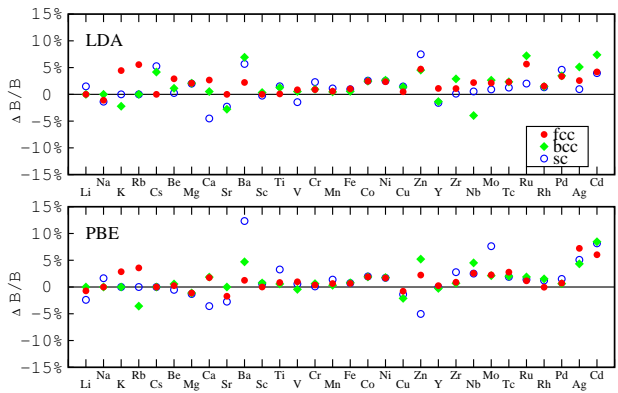}
\caption{Percentage differences between LAPW and PAW bulk moduli  
for the solids included in CMST, in all three crystalline phases considered,
calculated using LDA (top ) and PBE (bottom) functionals. 
}
\label{fig_B0}
\end{figure}

\begin{figure}
\subfigure[]{\includegraphics[width=0.99\columnwidth]{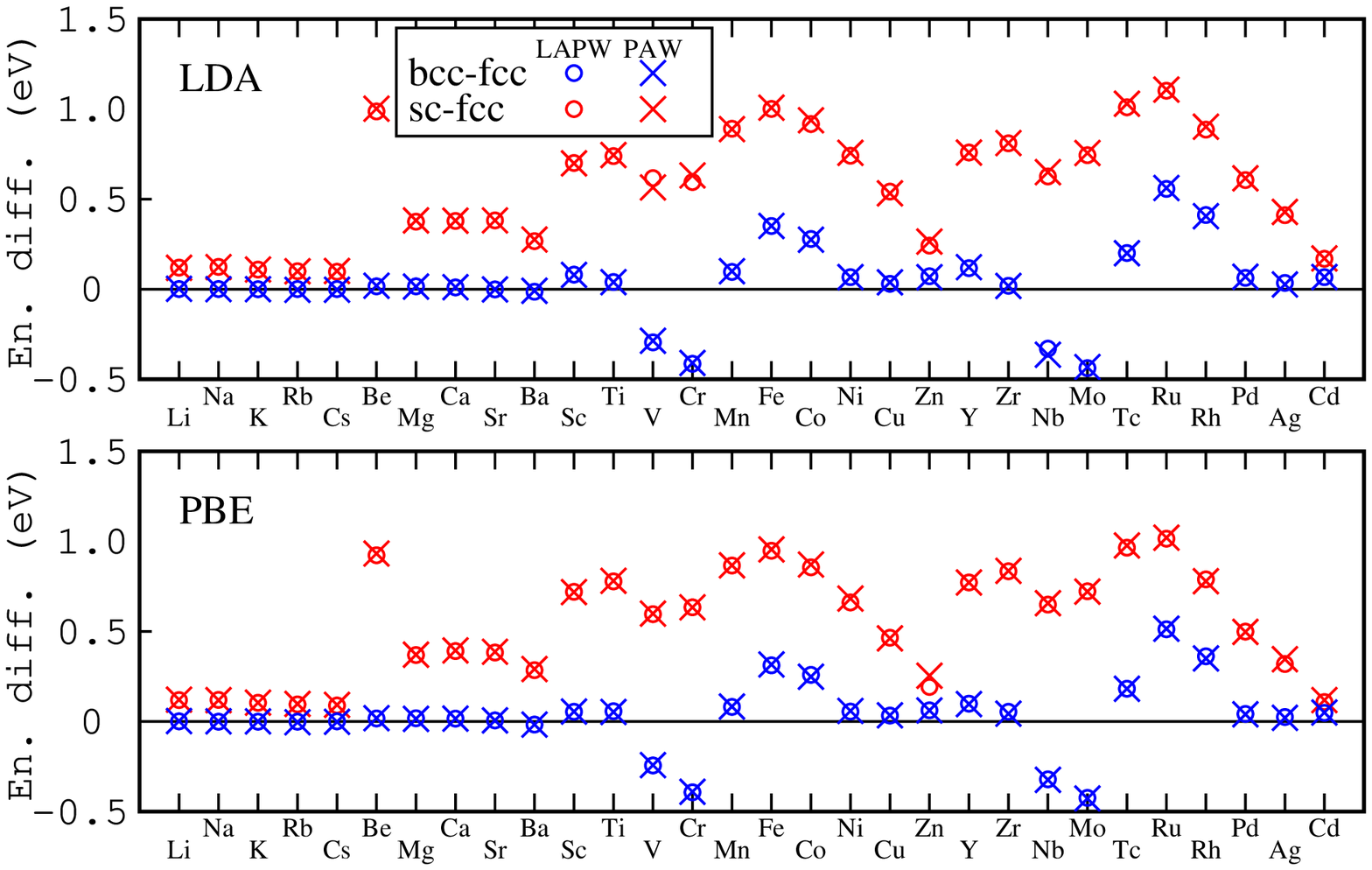}}
\subfigure[]{\includegraphics[width=0.99\columnwidth]{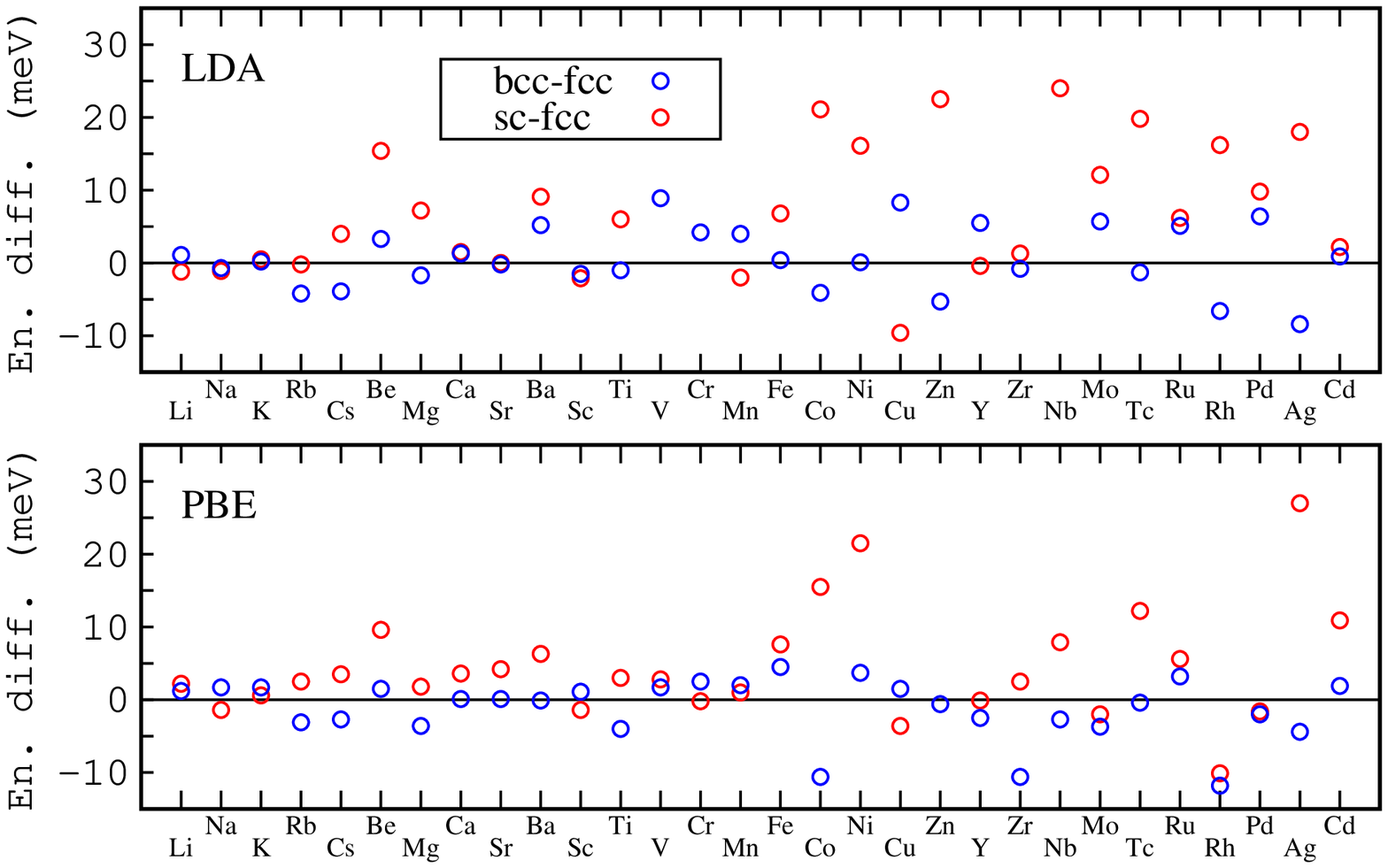}}
\caption{ (a) CMST absolute energy difference (eV) (with respect to \emph{fcc}) for the \emph{bcc} and \emph{sc} 
phases, for both FP-LAPW and PAW, 
using LDA (top) and PBE (bottom).
(b)Absolute differences (meV) between FP-LAPW and PAW estimates for the energy differences between the 
\emph{bcc-fcc} and \emph{sc-bcc} pairs, using LDA (top) and PBE (bottom).
}
\label{fig_e0}
\end{figure}

We report in Table~\ref{a0_B0_LDA} and \ref{a0_B0_PBE} the all-electron FP-LAPW
values for the equilibrium lattice constants ($a_0$), the 
bulk moduli ($B_0$) and the energy differences between the three
phases obtained using LDA and PBE respectively.
The energies of the \emph{bcc} and \emph{sc} structures are referenced to \emph{fcc} case. A negative value 
of the difference indicates that the structure is more stable than the \emph{fcc} one. 
Results for the PAW calculations are also reported in square brackets as differences with respect to 
the LAPW values. In Figs.~\ref{fig_a0} and \ref{fig_B0} the PAW-LAPW differences in $a_0$ and $B_0$ are plotted
as a percentage of the all-electron LAPW value. In Fig.~\ref{fig_e0} 
PAW and LAPW energy differences between phases are reported as well as the difference 
between PAW and LAPW results.

We start by discussing the differences in lattice constant
between PAW and LAPW calculations (Fig.~\ref{fig_a0}). 
For almost all elements and structures the PAW-LAPW discrepancy
on the lattice constant is well below 0.4\%. 
There are some notable exceptions: Ba (\emph{fcc} and \emph{bcc}),
and Cs (\emph{fcc}, \emph{bcc} and \emph{sc}), in LDA; Ba (\emph{fcc} and \emph{bcc}), Cs (\emph{sc})
and Rh (\emph{sc}) in PBE, with the worst cases being Cs \emph{sc} (-0.8\%) in LDA and Ba \emph{fcc} 
(-0.57\%) in PBE.
Most elements/structures show discrepancies smaller than 0.2\%: 78 and 75 out of 90 in LDA and PBE respectively.
Overall, the mean PAW-LAPW difference is very small both in LDA and PBE ( -0.065\%
and -0.055\% respectively).
The discrepancies between PAW and FP-LAPW in the \emph{fcc}, \emph{bcc}, and \emph{sc} lattice constants are 
not independent from each other as can be seen from the linear correlation coefficients 
in Table~\ref{rtable}. This high degree of correlation is quite important as it suggests 
that the pseudopotentials employed have excellent transferability among the electronic environments tested. 

\begin{table}[htb!]
\centering
\renewcommand{\arraystretch}{1.3}
\begin{tabular}{@{}lrcrcr@{}}
\toprule
       & $fcc$ & \phantom{abc} & $bcc$ & \phantom{abc} & $sc$ \\
\midrule
$fcc$ & 1     && 0.92  && 0.86   \\
$bcc$ & 0.94  && 1     && 0.89   \\
$sc$  & 0.90  && 0.88  && 1    \\
\bottomrule
\end{tabular}
\caption{Correlation matrix between PAW and FP-LAPW differences for the equilibrium lattice constant
$a_0$ in the \emph{fcc}, \emph{bcc}, and \emph{sc}
structures. The upper half of the matrix shows LDA results while the 
lower half of the matrix shows PBE results.
A correlation matrix element close to 1 indicates a strong correlation, 
highlighting that the performance of the 
pseudopotentials used in this work is similar in the different electronic environments considered.}
\label{rtable}
\end{table}

Overall, from the data collected it appears that the comparison between PAW and FP-LAPW
is slightly closer for PBE than for LDA. Also, we find that for some elements the treatment of 
the semicore electrons has a significant effect, in particular
for Cs and Ba the inclusion of the  $4d$ electrons in the core would
result in smaller lattice parameters by as much as 0.03 \AA~in the LAPW calculations. 
Due to this effect here we report \texttt{Elk} results with $4d$ electrons in valence for the case of Ba and Cs.
However, the impact of semicore states appears less important in the PAW calculations; 
therefore the results reported use PAW datasets with $4d$ electrons in the core.

Similar PAW calculations have been reported in Ref.~\onlinecite{GBRV};
however, in that study a fixed cut-off of $40$ Ry has been used for all 
elements to optimize sets for high-throughput calculations. 
Although for many elements there is agreement between our data and those of 
Ref.~\onlinecite{GBRV}, in some cases, such as Li and Cu, the
relatively large deviation reported in that work is not confirmed by
our calculations. We attribute this difference to the use of insufficient number of 
planewaves in Ref.~\onlinecite{GBRV}. 
On the contrary, in the case of Cd both our calculations
and the ones in Ref.~\onlinecite{GBRV} consistently show a relatively large deviation
from all-electron results.

The PAW~-~FP-LAPW deviations for the bulk moduli are reported in Fig.~\ref{fig_B0}.
For most of the elements and structures studied, the PAW~-~FP-LAPW difference is less
than 6\%. The exceptions are
 K (\emph{fcc}), Rb (\emph{fcc}), Cs (\emph{bcc}), Ba (\emph{bcc} and \emph{sc}), Zn (\emph{sc}),
 Ru (\emph{fcc} and \emph{bcc}), Ag (\emph{bcc}),
 and  Cd (\emph{bcc}) in LDA; Ba (\emph{sc}), Zn (\emph{bcc} and \emph{sc}), Mo (\emph{sc}),
 Ag (\emph{fcc} and \emph{sc}),
 and Cd (\emph{fcc}, \emph{bcc} and \emph{sc}) in PBE. 
The average deviation between PAW and LAPW bulk moduli is 1.7\% for LDA and 
1.3\% for PBE, with a general trend of PAW overestimating FP-LAPW $B_0$.

Our calculations reveal that, contrary to common expectation, 
the percentile deviation on $B_0$ is not tightly correlated with the
corresponding deviation on $a_0$: the linear correlation coefficient
between these two data sets can be as low as -0.26 for LDA and -0.31
for PBE.

We report the PAW~-~FP-LAPW deviation for the $bcc$-$fcc$
and $sc$-$fcc$ energy differences in Fig.~\ref{fig_e0} 
While we require an accuracy of the order of a few mRy in the transferability
tests of the PPs, this error is usually an upper limit and for many elements 
the energy differences between different phases agree with the all-electron results within a few meV. 
A deviation larger than 13 meV ($\approx$1 mRy) in the $bcc$-$fcc$ energy difference is found
only for Nb in LDA. The arithmetic average of the PAW~-~FP-LAPW deviations are as small as 2~meV and -1~meV in 
LDA and PBE, with a standard deviation of 14~meV and 4~meV respectively. 
The PAW~-~FP-LAPW discrepancies in the $sc$-$fcc$ energy difference are somewhat larger 
than the ones for $bcc$-$fcc$ energy difference, consistent with the larger average 
value of these differences.
Choosing a threshold of 26 meV ($\approx$ 2 mRy) one can identify V and Cr as the elements 
needing further attention in LDA and Zn and Ag in PBE.
The arithmetic average of the PAW~-~FP-LAPW deviations are 9~meV and 7~meV for LDA and PBE, with 19~meV and 
6~meV standard deviation respectively.
The $sc$-$fcc$ PAW~-~FP-LAPW deviations appear to be positive biased, while
$bcc$-$fcc$ are centered around zero. 
Overall, in the case of the widely used PBE functional, an excellent agreement 
between our PAW datasets and LAPW results can be seen. 

Results for alkaline-metal and alkaline-earth elements deserve a word of caution.
Our FP-LAPW calculations have been converged within 1 meV/atom and the 
transferability of the \texttt{QE-PAW} datasets was verified to be of a few mRy in a number of atomic configurations.
As the energy difference between \emph{fcc} and \emph{bcc} structures in
alkaline and alkaline-earth metals are of the order of 1 meV
and 1 mRy respectively, some difficulty in accurately reproducing these 
within the present study can be expected. 
Nevertheless, general trends seem in agreement with literature.
As an example, in Li, K, and Rb the $fcc$ structure is known to be more stable than the $bcc$ when using LDA.
\cite{Sliwko1996,Nobel1992}
We confirm these results both for LAPW and PAW, except for Rb within PAW which is stable in the $bcc$ structure.
For alkaline-earth elements LAPW and PAW calculations show identical outcomes, 
identifying correctly the experimental equilibrium phase for Ca ($fcc$), and Ba ($bcc$).
The equilibrium phase in Be and Mg is the hexagonal-close packed ($hcp$) structure not included in our test set; 
still, we notice that the close-packed $fcc$ phase is favored with respect to $bcc$ in these two elements, as 
could be expected.
The \emph{bcc}-\emph{fcc} energy difference in Sr is very small, just a few meV, and only PBE 
(both within PAW and LAPW) captures the correct ($fcc$) ground state phase.

\begin{figure}
\includegraphics[width=0.5\textwidth]{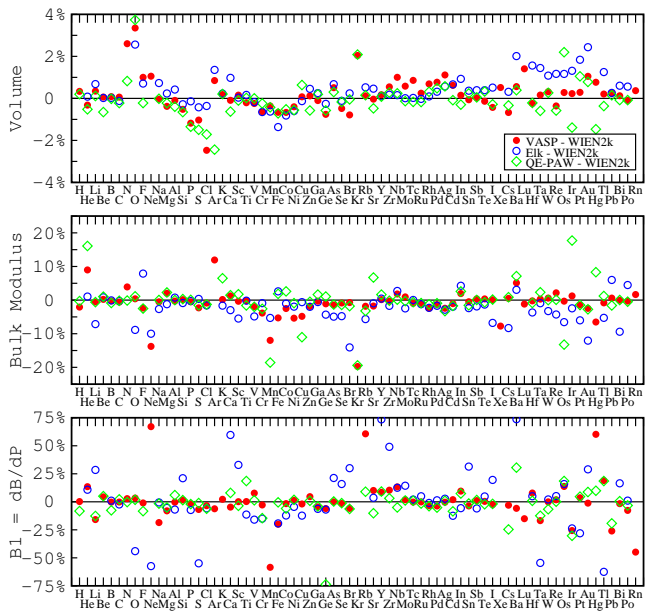}
\caption{Comparison of EoS parameters between \texttt{QE-PAW} dataset, 
\texttt{VASP} PAW, FP-LAPW results from \texttt{Elk} and \texttt{WIEN2k}. The zero reference is chosen as 
\texttt{WIEN2k}
for all elements considered in the $\Delta$ factor calculations. 
All calculations are performed using PBE exchange-correlation functional;
\texttt{VASP} and \texttt{WIEN2k} values taken from Ref.~\onlinecite{Lejaeghere2013}.}
\label{fig_Etot}
\end{figure}
\begin{figure*}
\includegraphics[width=0.49\textwidth]{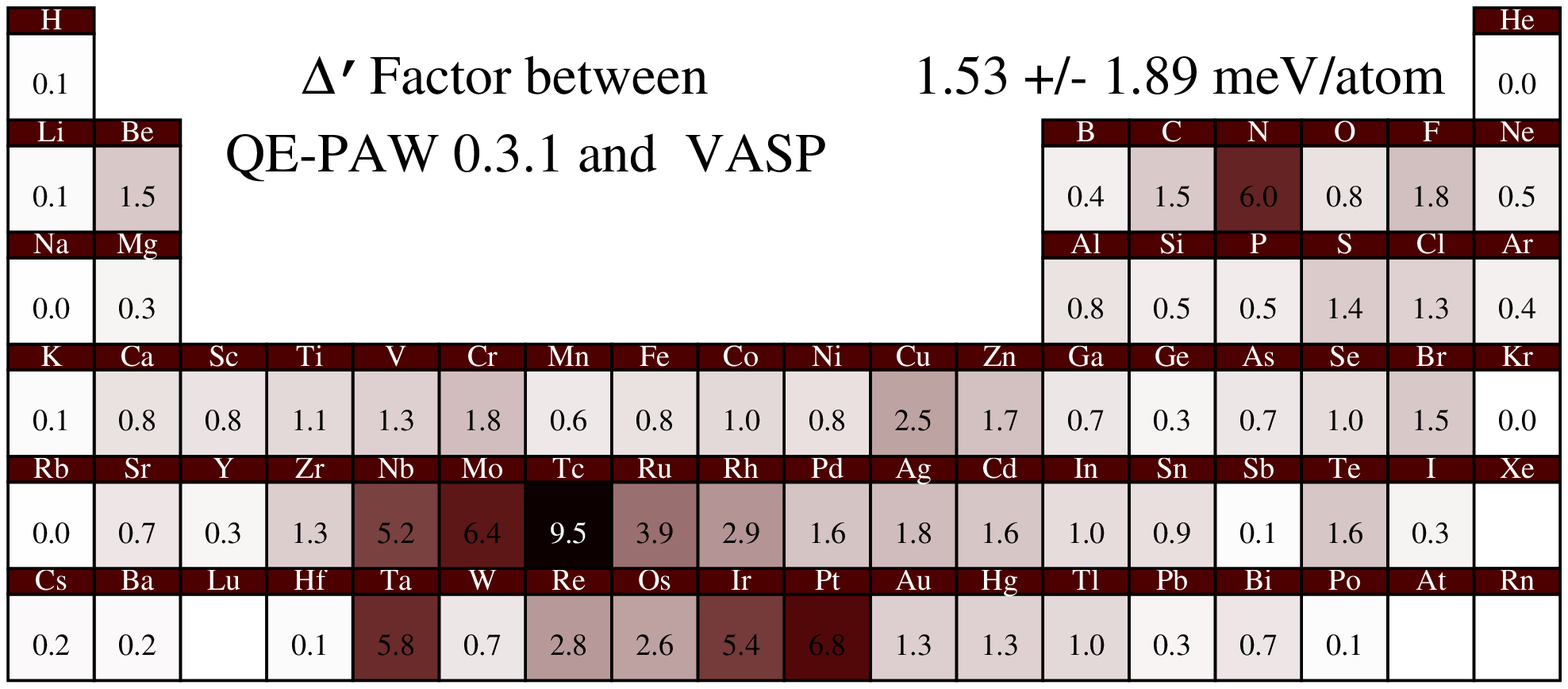}
\includegraphics[width=0.49\textwidth]{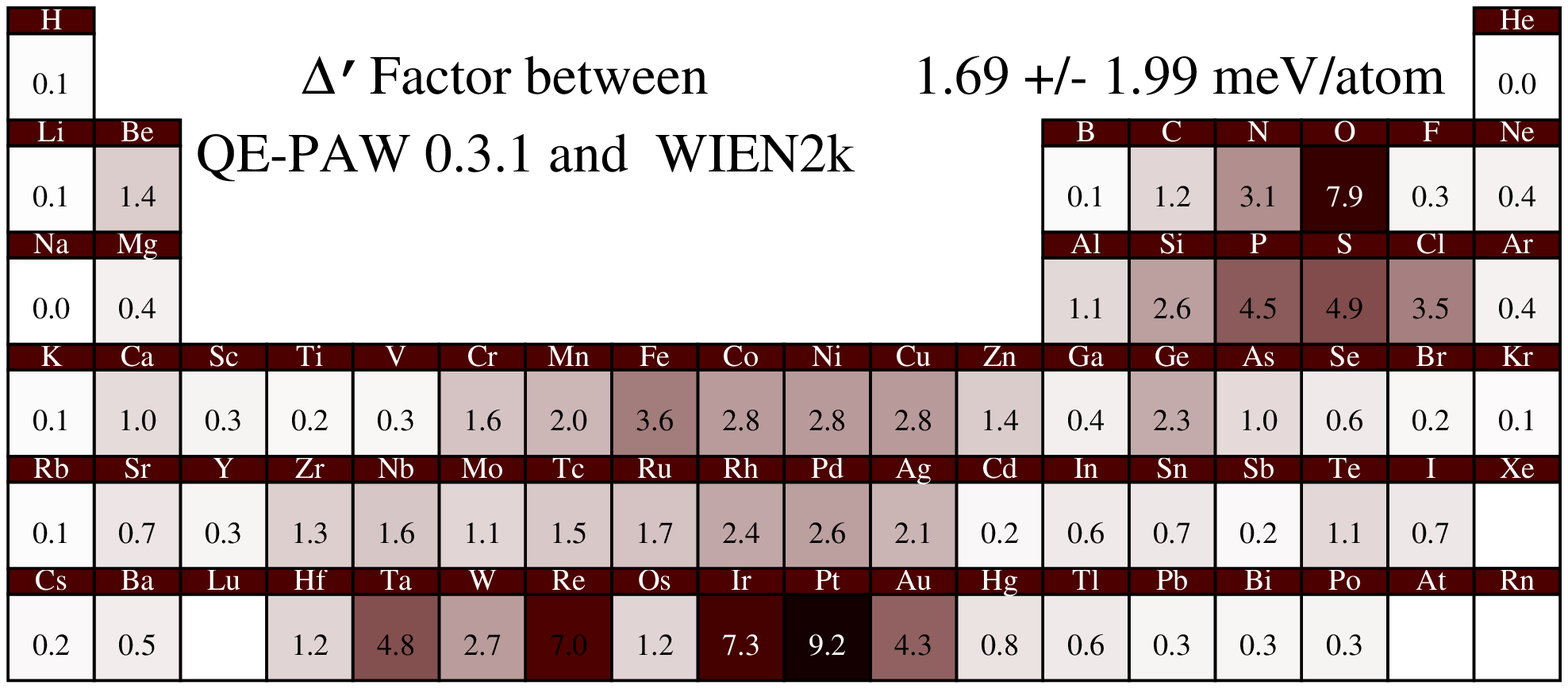}
\includegraphics[width=0.49\textwidth]{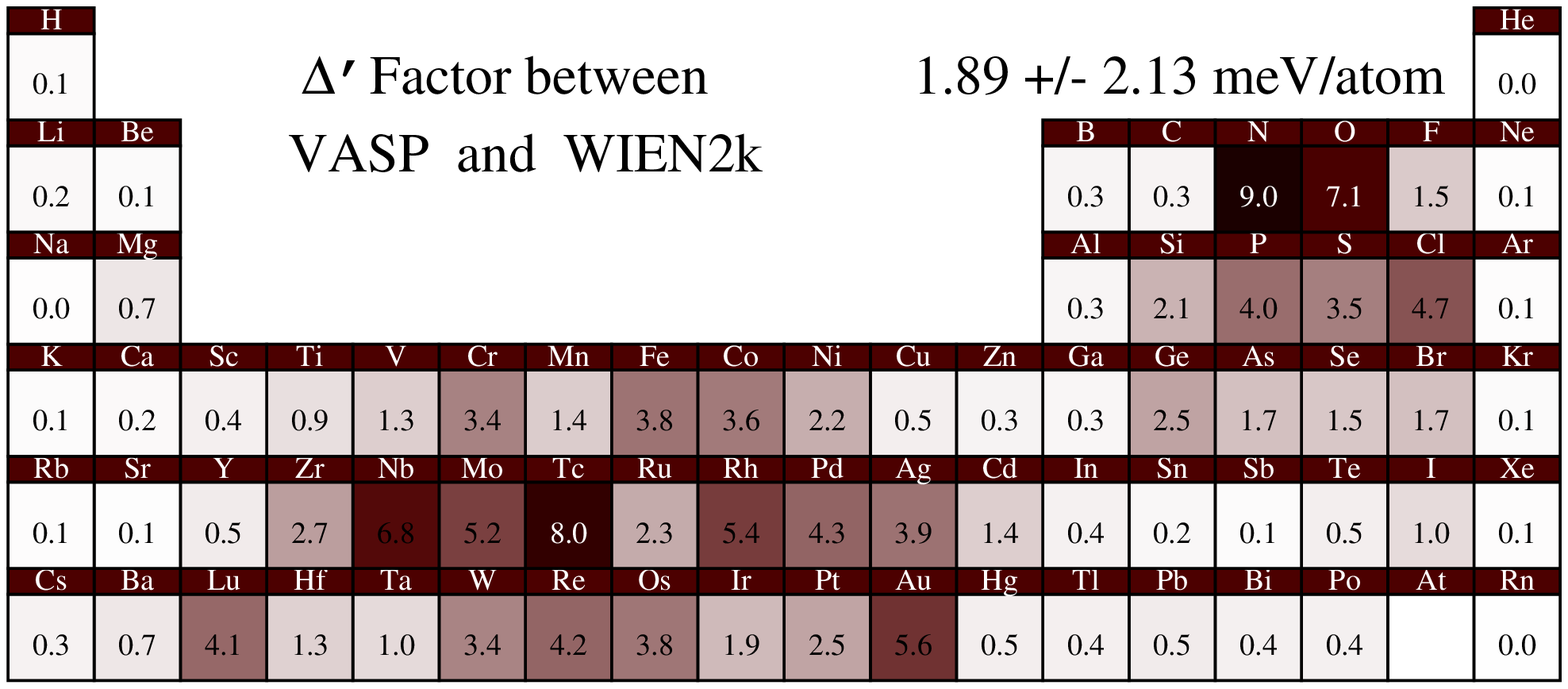}
\includegraphics[width=0.49\textwidth]{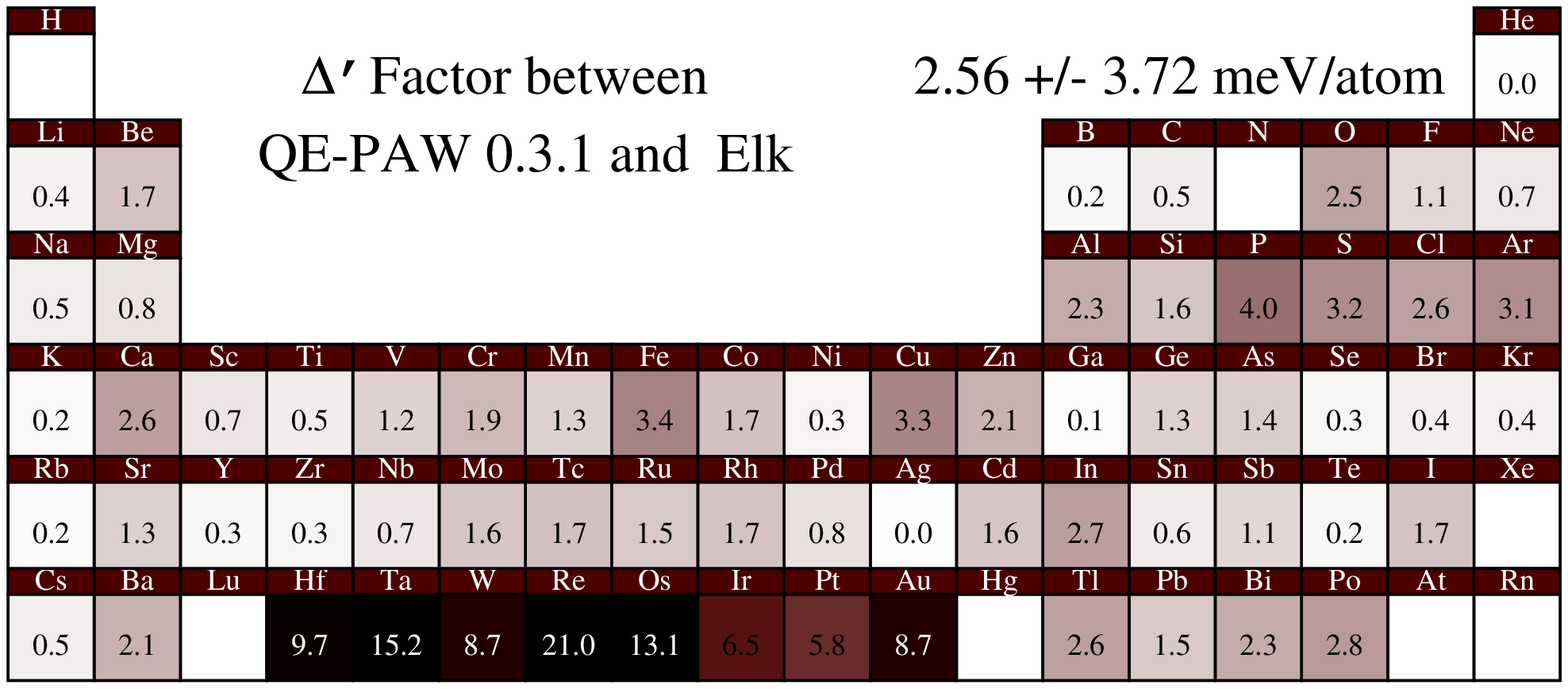}
\includegraphics[width=0.49\textwidth]{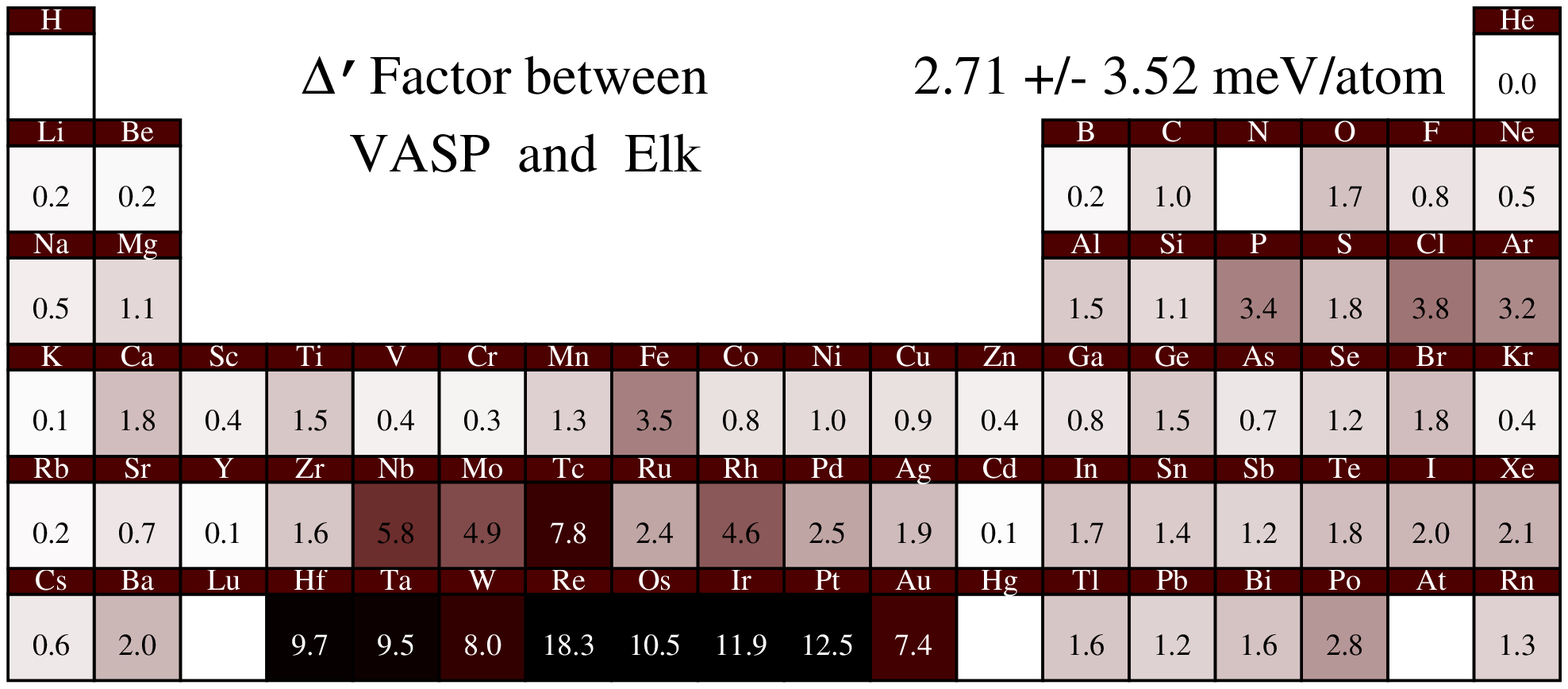}
\includegraphics[width=0.49\textwidth]{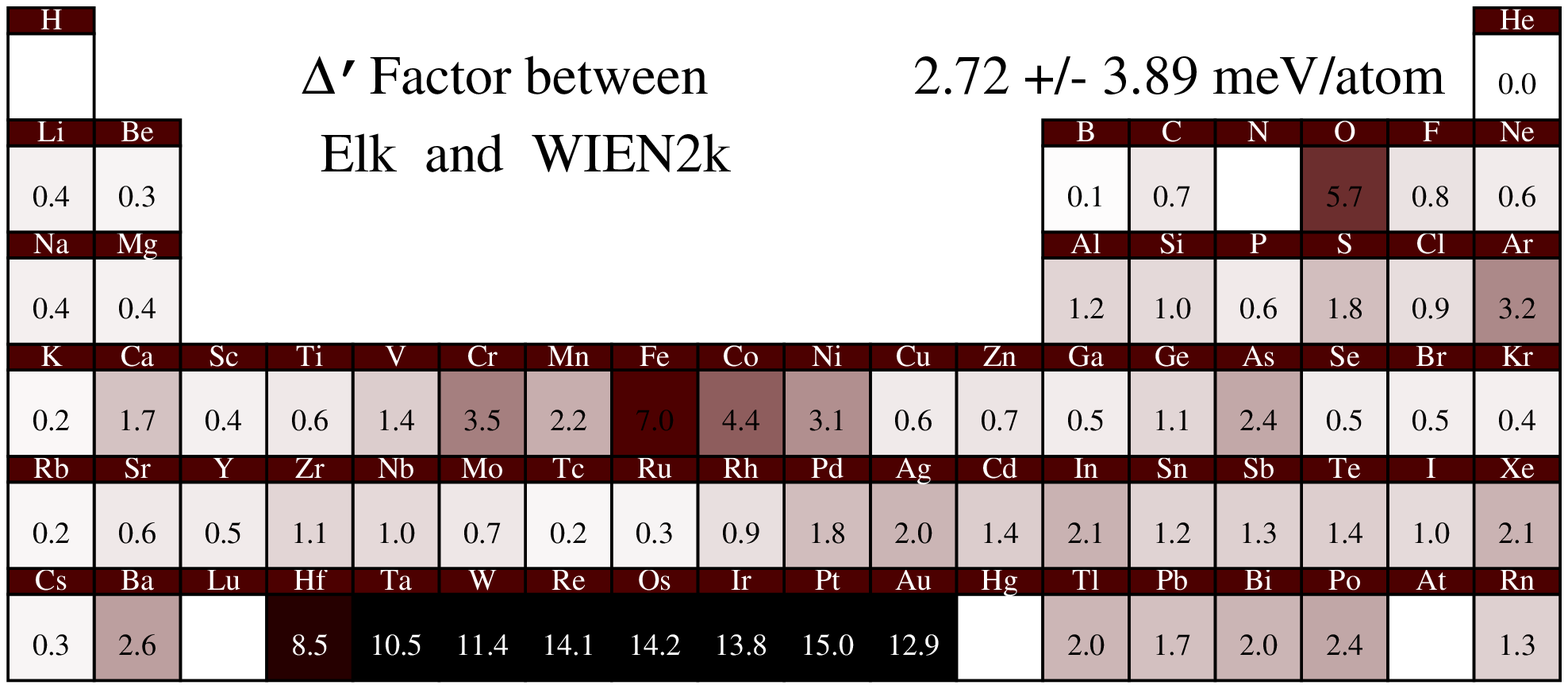}
\caption{$\Delta^{\prime}$ factor comparison between \texttt{QE-PAW}, 
\texttt{VASP} PAW and FP-LAPW results from \texttt{Elk} and \texttt{WIEN2k}. $\Delta^{\prime}$
is a symmetric measure that qualifies as a distance between the codes/methods considered 
as it obeys the triangular inequality criterion. For each pair we 
provide the average distance as well as individual distances per element. Color codes are linear 
with respect to distances. 
All calculations are performed with the PBE exchange correlation functional.
\texttt{VASP} and \texttt{WIEN2k} values are taken from Ref.~\onlinecite{Lejaeghere2013}}
\label{fig_Delta}
\end{figure*}

\begin{figure*}
\includegraphics[width=0.49\textwidth]{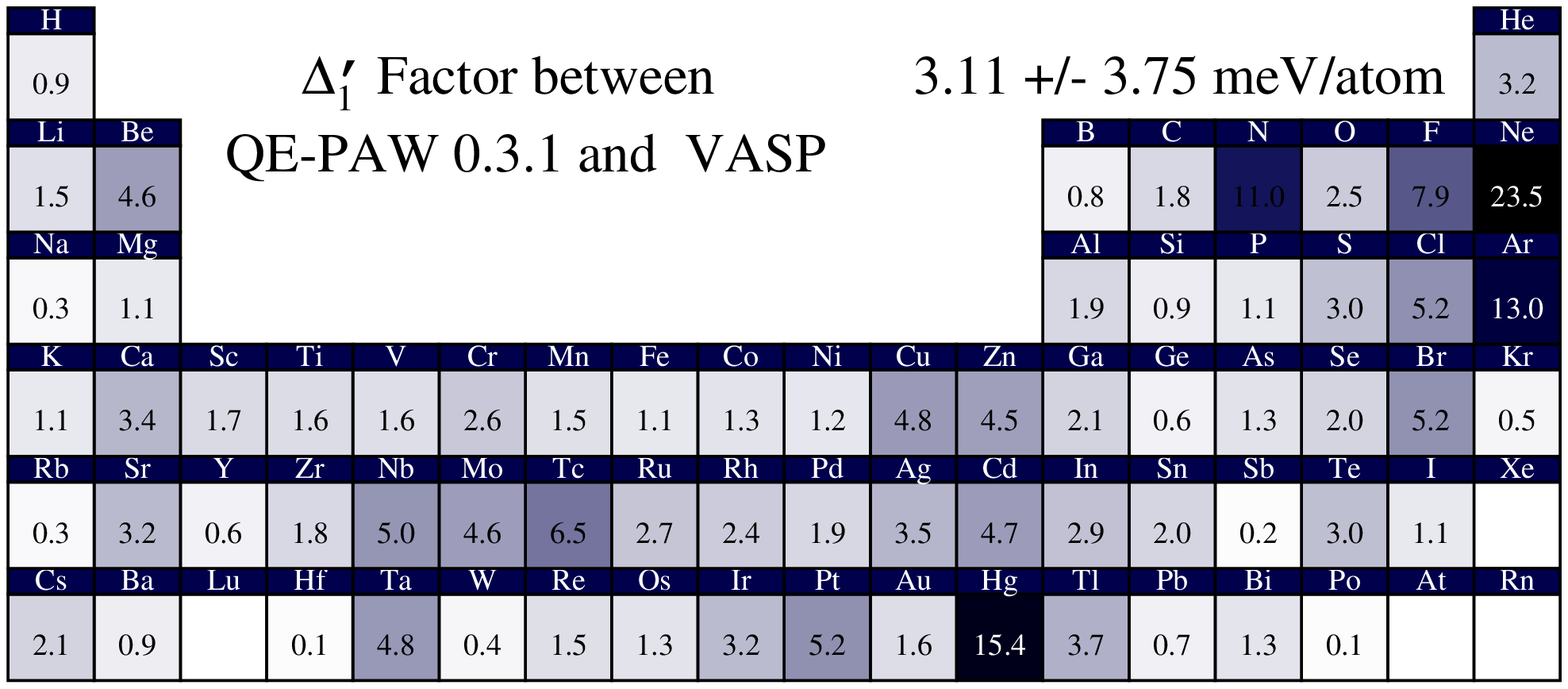}
\includegraphics[width=0.49\textwidth]{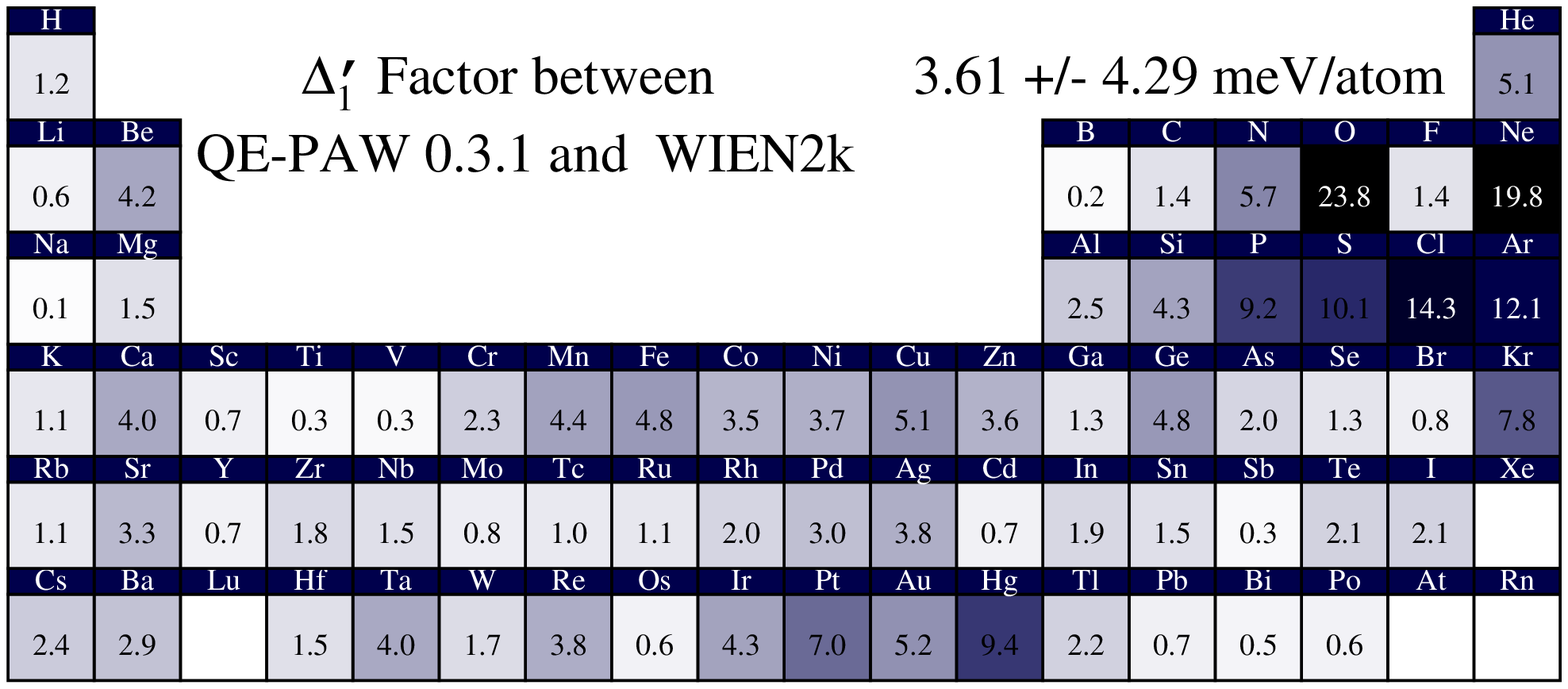}
\includegraphics[width=0.49\textwidth]{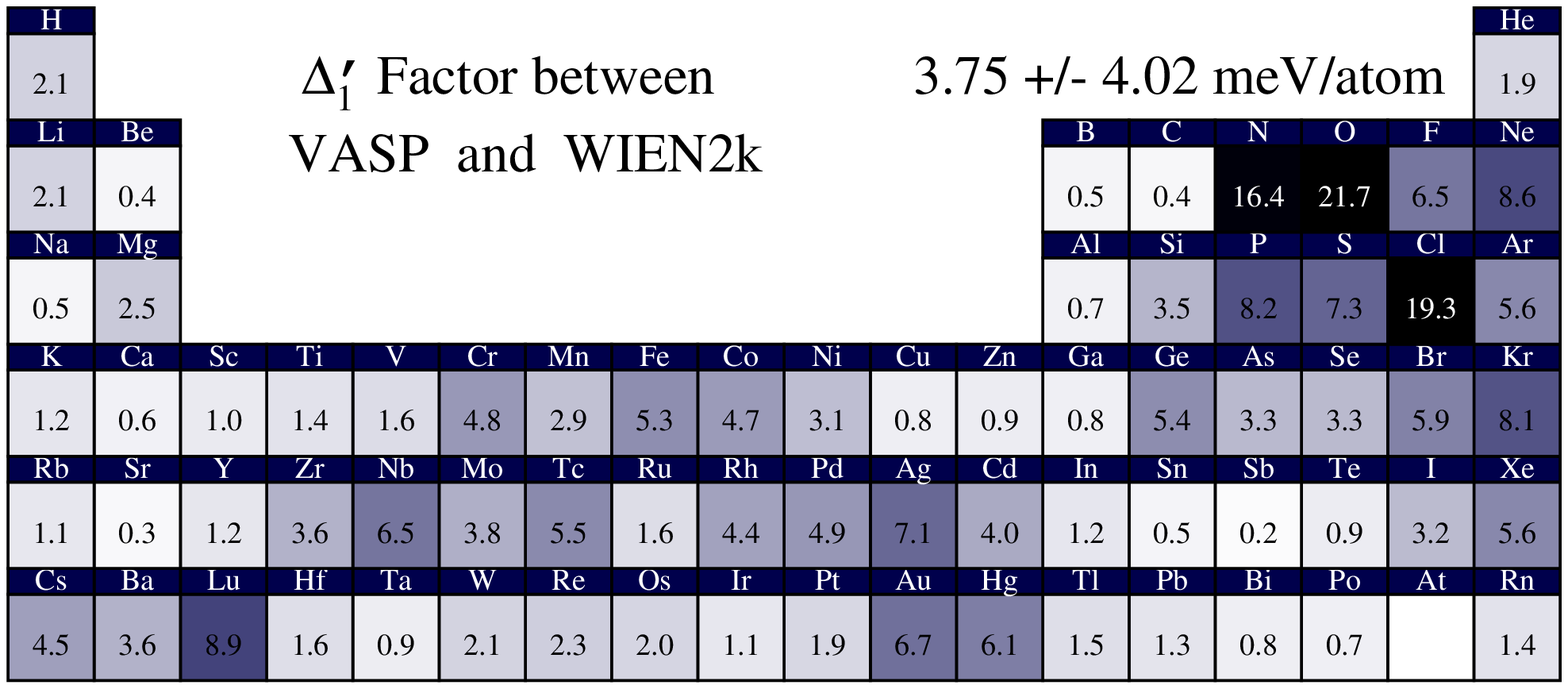}
\includegraphics[width=0.49\textwidth]{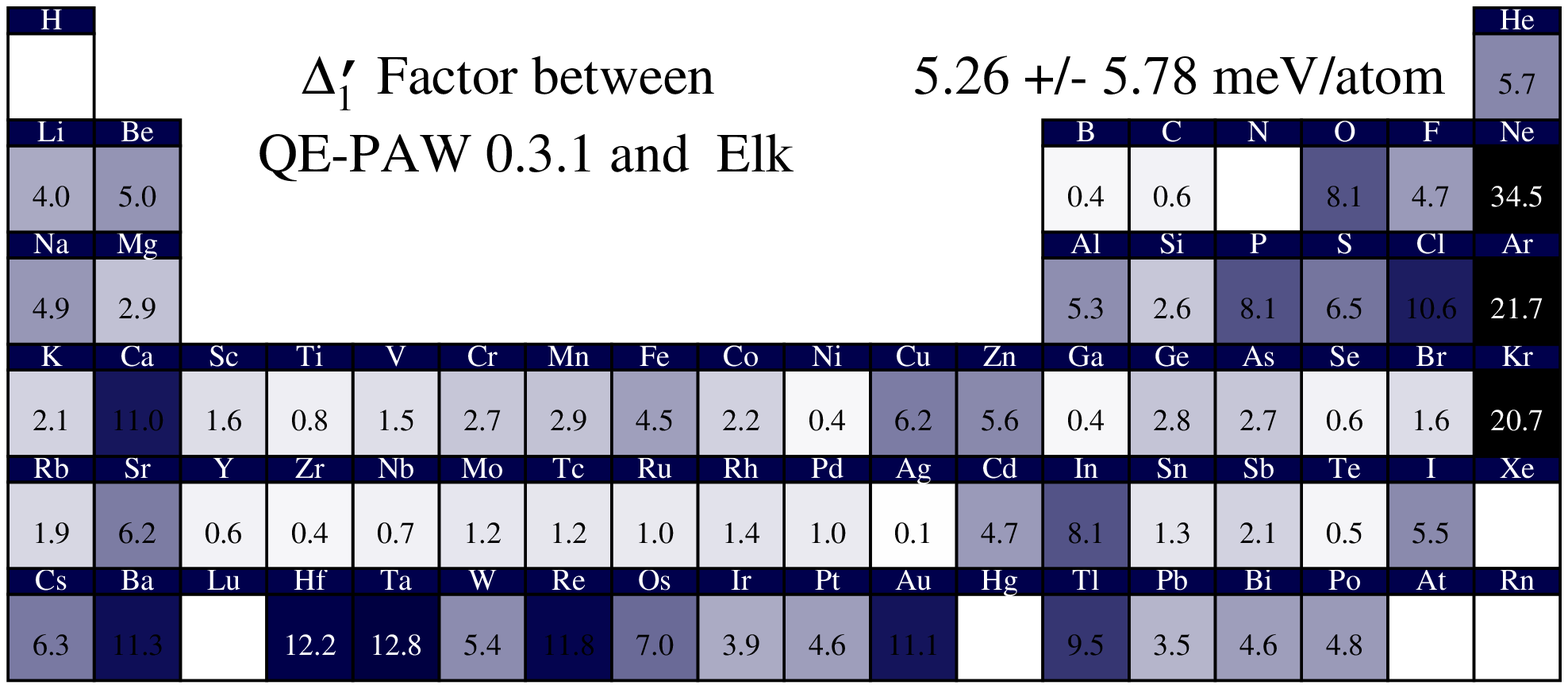}
\includegraphics[width=0.49\textwidth]{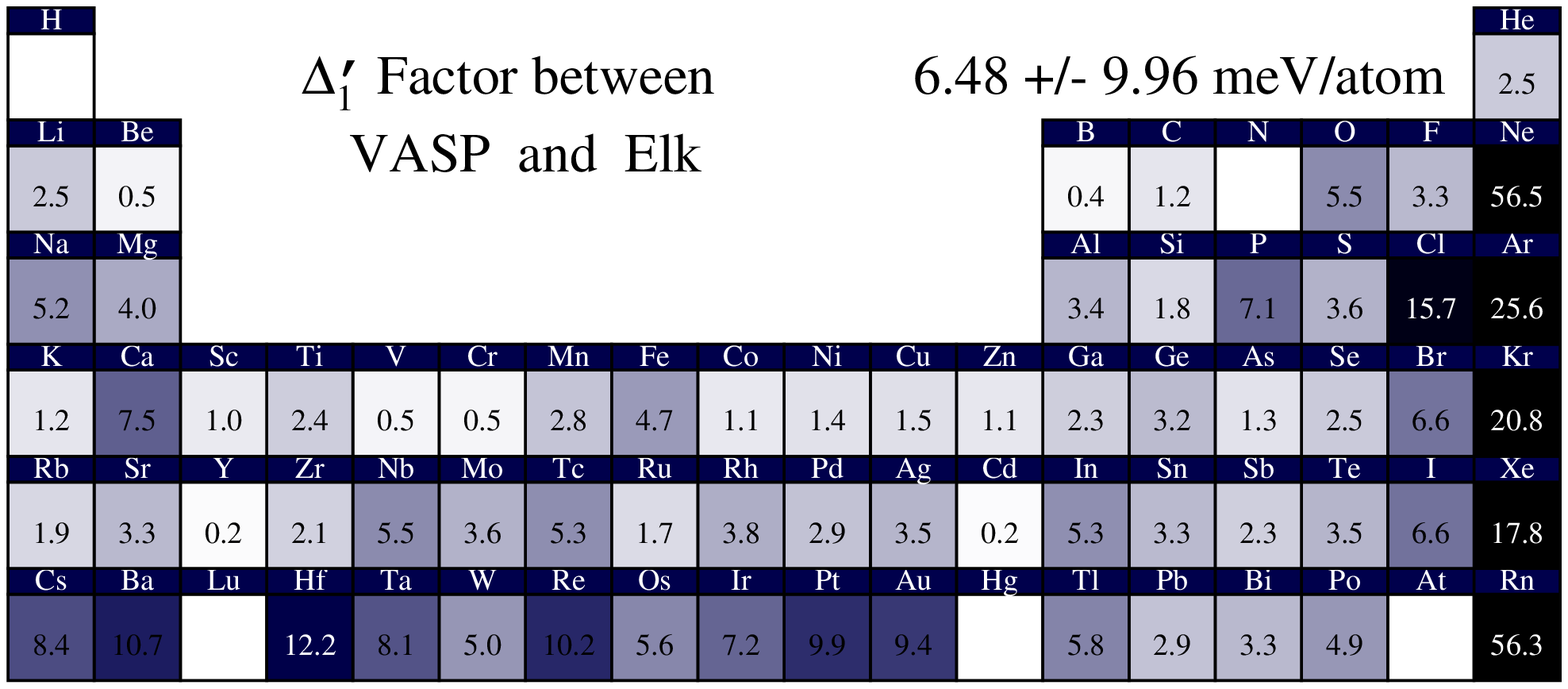}
\includegraphics[width=0.49\textwidth]{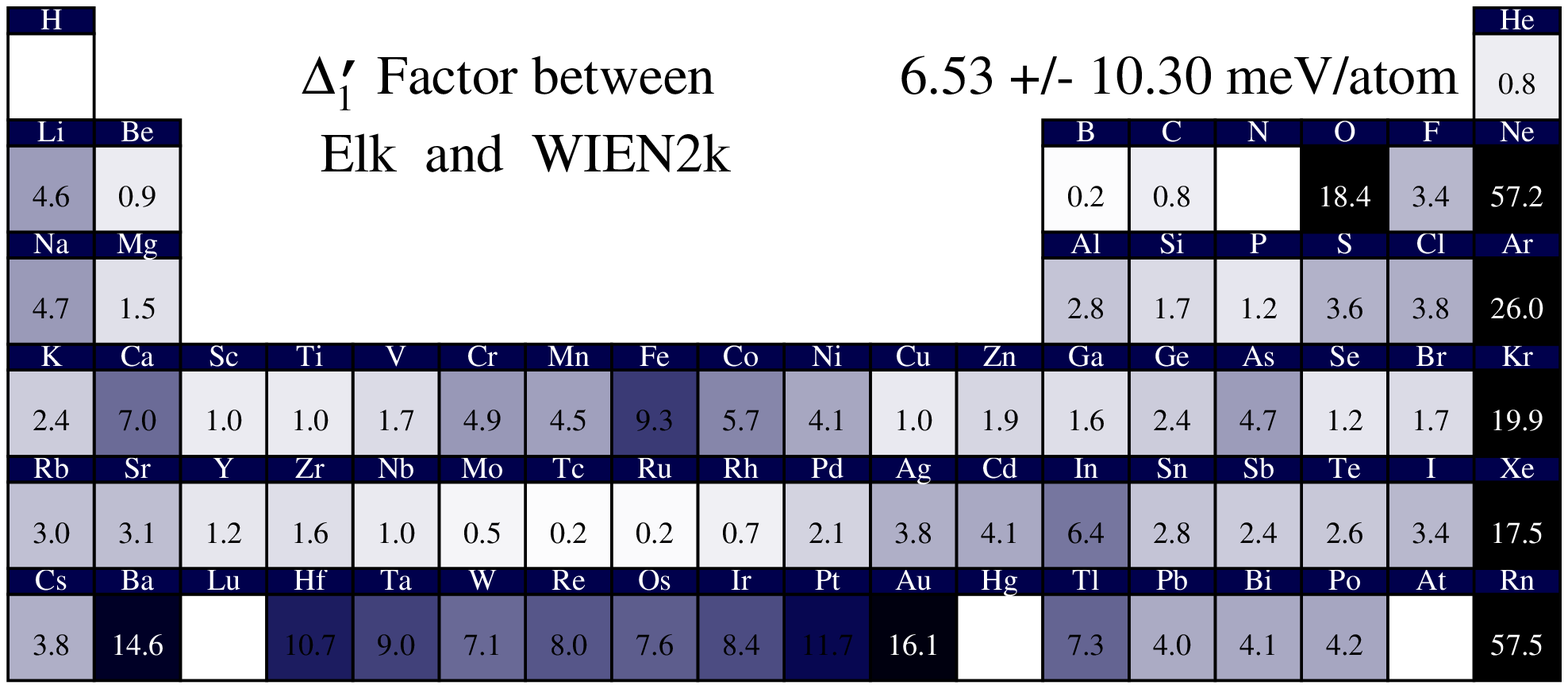}
\caption{ $\Delta_1^{\prime}$ factor comparison between \texttt{QE-PAW}, 
\texttt{VASP} PAW and FP-LAPW results from \texttt{Elk} and \texttt{WIEN2k}. $\Delta_1^{\prime}$
is a symmetric and positive-definite measure and does not qualify as a distance.
However, this measure removes the bias of the original $\Delta$ factor with respect to 
stiffness and therefore yields a criteria to assess differences between codes/methods
that is more homogeneous over all periodic table.
For each pair we provide the average $\Delta_1^{\prime}$ and the standard deviation, 
as well as the individual distance for each element. 
Color codes are linear with respect to $\Delta_1^{\prime}$ factor.
All calculations are performed with the PBE exchange correlation functional.
\texttt{VASP} and \texttt{WIEN2k} values are taken from Ref.~\onlinecite{Lejaeghere2013}
}
\label{fig_Delta1}
\end{figure*}

\subsection{$\Delta$ Factor across the periodic table}
\label{paw_test_emine}

In this section we present the results of the calculation of the
structural properties of a large majority of the elements
in the periodic table (68) in their experimental equilibrium phase, (or occasionally a simpler, closely related one),
obtained following the procedure described in Ref.~\onlinecite{Lejaeghere2013}.

Results obtained by four codes are compared: 
i) \texttt{Elk} LAPW calculations performed following the protocol described in section~\ref{AE_tech}  
ii) \texttt{QE-PAW} calculations as described in section~\ref{PAW_tech} 
iii) \texttt{VASP}  
and iv)  \texttt{WIEN2k} results from the literature.
The data for \texttt{VASP} and \texttt{WIEN2k} have been determined in 
Ref.~\onlinecite{Lejaeghere2013} and made available online at the CMM
website \cite{CMM-website}, and the most recent values available at the website are used here.\cite{recent} 
In our \texttt{QE-PAW} calculations results for all elements
discussed in Ref.~\onlinecite{Lejaeghere2013} are presented, with the
exception of Lu, Rn and Xe whose datasets are still under development.
\texttt{Elk} calculations for H, N, Lu and Hg are also excluded due to convergence problems.

In Fig.~\ref{fig_Etot} we display the relative deviation from the
\texttt{WIEN2k} results of the structural parameters $V_0$, $B_0$ and $B_0'$,
obtained  fitting Eq.~\ref{eq:E(V)} to the 
\texttt{Elk} and \texttt{QE-PAW} calculations. For completeness, 
the \texttt{VASP} results are also included. 
All numerical values of all these parameters are given 
in full in the Supplementary Material.

We can see that values predicted by different computational methods
show a different spread depending on the property considered. The equilibrium volume
has a typical variation of 1-2\%, although larger deviations are
occasionally present, while the bulk modulus has a larger spread of 5-10\%, 
and its pressure derivative an even larger one.

In order to quantify in a single figure the agreement between the results
of different codes Lejaeghere and coworkers\cite{Lejaeghere2013} introduced a quality factor $\Delta$, measuring 
the discrepancy between the corresponding two EoS (see section~\ref{sec_delta} above).
In the original definition of $\Delta$ the
volume integration is defined as a fixed $\pm$6\% interval around the
equilibrium volume obtained for one of the two methods (\texttt{WIEN2k} in their case) 
that is taken as a reference.  
This approach  
cannot be completely satisfactory 
because it results in an asymmetric comparison,
since it depends on the choice of a reference data set among the two elements of each comparison. 
Instead, having access to a measure of the {\it distance} between the results of two codes/methods
rather than taking one of the two as an absolute reference could provide further insight 
in the comparison as the number of codes/methods to compare increases.

We tried to correct for this by defining a symmetric quality factor $\Delta_{sym}$
where the volume integration in Eq.~\ref{eq:delta} is performed in the $\pm$6\% interval 
around the {\it average} equilibrium volumes predicted by the two methods being compared. 
We note that $\Delta_{sym}$, although being a symmetric and positive definite function, fails to satisfy 
the triangular inequality for the distances between any three codes/methods, also explicitly 
checked on the available data.

We therefore introduce a slightly modified quality factor $\Delta'$ 
defined so that the volume integral is performed in the $\pm$6\% interval around
the {\it reference} volume $V_{ref}$ used to generate the 7 points in the energy-volume curve 
which are used to determine the EoS. The resulting $\Delta'$
defines a proper distance between pairs of codes/methods for each element.
Very satisfactorily, the computed $\Delta'$ values differ only slightly from the original $\Delta$ values, while
being a well defined distance between codes and methods.

In Fig.~\ref{fig_Delta} we display the results of the $\Delta'$ factors
across the periodic table comparing \texttt{Elk},
\texttt{VASP}, \texttt{WIEN2k}, and \texttt{QE-PAW}.
On average the four codes/methods agree very well with each other.
The two that are closest are \texttt{QE-PAW} and \texttt{VASP} (1.53 meV/atom), 
\texttt{WIEN2k} is only marginally more distant
(1.69-1.89 meV/atom) and  \texttt{Elk} is slightly farther away (2.56-2.72 meV/atom) due to the contributions of the $5d$
transition metal elements. This discrepancy between two all-electron calculations
highlights the importance of standardized verification and validation tests,
and indicates that optimization of computational parameters 
plays an important role in the outcome of the FP-LAPW calculations.
Other elements for which noticeable discrepancies
between methods are visible are certain transition metals and the second half 
of the elements of the first row, 
revealing once more the elements which require careful treatment within DFT, 
where small changes in implementation can have a significant impact in the outcome.
Besides these cases, all methods give consistently very close results.

It has been pointed out in Ref.~\onlinecite{JTH} that the value of $\Delta$ factor
strongly depends on the stiffness of the material and on its volume per atom.
The larger the bulk modulus or the atomic volume of a given element is, the 
larger the $\Delta$ value associated to a given deviation of the structural parameters becomes.

A ``renormalized'' $\Delta_1$ factor has been proposed\cite{JTH},
\begin{equation}
 \Delta_1 = \frac{V_{ref}B_{ref}}{V_0 B_0} \Delta, 
 \label{eq:delta1}
\end{equation}
where the original value of $\Delta$ is scaled by the ratio of the
equilibrium volume and the bulk modulus with respect to some reference
values, taken to be $V_{ref} = 30$ \AA,  $B_{ref} = 100 $ GPa, that 
roughly correspond to their average values over the elements.

The resulting $\Delta_1$ factor gives a more homogeneous measure of
the quality of the agreement between codes across the periodic table.

We define a modified $\Delta_1'$ rescaling the previously defined
$\Delta'$ factor via Eq.\ref{eq:delta1} and taking for each element 
$V_0$ as the central
value of the volume integration interval and $B_0$ as the average of
the bulk moduli computed for the two codes/methods to be compared.
$\Delta_1'$ is no more a well-defined distance; however, it is imperative to analyze
the results with this quality factor to understand better the effects of stiffness 
and volume per atom on EoS comparisons. 

In Fig.~\ref{fig_Delta1} we display the result for the $\Delta_1'$ factor calculations
across the periodic table for the four codes/methods.
On average $\Delta_1'$ values are doubled with respect to $\Delta'$.

\begin{table}[ht!]
\centering
\begin{tabular}{ll|cc|cc}
~ & ~ & $\Delta$ & $\Delta'$ & $\Delta_1$ & $\Delta_1'$  \\
\hline
 \texttt{WIEN2k} & \texttt{QE}            &~ 1.70 (1.70) & ~ 1.69 &~ ~4.18 ~(3.46)& ~3.61 \\
 \texttt{WIEN2k} & \texttt{VASP}          &~ 1.92 (1.89) & ~ 1.89 &~ ~3.79 ~(3.78)& ~3.75 \\
 \texttt{WIEN2k} & \texttt{Elk}  &~ 2.73 (2.71) & ~ 2.72 &~ 13.27 ~(5.55)& ~6.53 \\
 \texttt{Elk} & \texttt{QE}      &~ 2.58 (2.64) & ~ 2.56 &~ ~5.78 ~(6.96)& ~5.26 \\
 \texttt{Elk} & \texttt{VASP}    &~ 2.72 (2.71) & ~ 2.71 &~ ~5.33 (11.72)& ~6.48 \\
 \texttt{QE} & \texttt{VASP}              &~ 1.56 (1.53) & ~ 1.53 &~ ~2.97 ~(3.79)& ~3.11 \\
\end{tabular}
\caption{PBE results for various $\Delta$ factor definitions in meV. 
For $\Delta$ and $\Delta_1$ factors, which are not symmetric,
in parenthesis we give the reverse comparison result where the 
second method is chosen as the reference instead.
}
\label{table:summary}
\end{table}

Again, the two codes/methods that result to be closest are \texttt{QE-PAW} 
and \texttt{VASP} (3.11 meV/atom). 
\texttt{WIEN2k} is only marginally more distant from each of them,
(3.61-3.75 meV/atom) while \texttt{Elk} is farther away (5.25-6.53 meV/atom).
Examining the data one can confirm that the second half of the 
elements of the first row as well as elements 
in the transition metal series remain problematic, although $5d$ elements appear less 
so here than when considering the $\Delta'$ factor.
Another class of 
potentially challenging elements are the 
noble gases, for which the values of the original $\Delta'$ factor were 
systematically very small due to the very low value of their bulk moduli.
We have thus re-examined further the 
EoS for the noble gases. It is found that the 
energy differences in EoS are within the convergence limit of our calculations
for Ar, Kr, Xe in \texttt{Elk} and Ne in \texttt{QE-PAW}.
These findings reveal that the bias of the original $\Delta'$ factor with respect bulk moduli 
can be overcome with the use of $\Delta_1'$.

In Table~\ref{table:summary} we provide a final summary of our results using all
variations of $\Delta$ factors that have been proposed so far. 
Our results show that for all the variations in the $\Delta'$ factor, 
all the compared codes/methods agree well within 15 meV. 
We also see that rather than an average over the periodic table, 
an element-by-element analysis of the quality factor, together with the EoS when necessary,
highlights the elements that could benefit from improvement in each computational approach.
Following these indications our efforts towards more standardized pseudopotentials 
continue (see the experimental pseudization recipes in \texttt{PSLibrary v1.0.0} \cite{dalcorso100}). 

\section{Conclusions}

\label{conclusions}
In this study we have introduced a Crystalline Monoatomic Solid Test protocol comprising 
the calculation of the structural properties and relative energy
differences of the three crystalline non-magnetic cubic phases ($sc$, $fcc$, and
$bcc$) of any given element.
While most results in the literature focus on experimentally stable phases, 
these results provide an extensive comparison of the energetics
of simple but realistic structural phases that explore several coordination numbers, 
thus providing key information to assess and compare the performance of different
computational methodologies.

We have collected a database of carefully performed all-electron FP-LAPW results, obtained by
the \texttt{Elk} open-source code, and plane-wave PAW results,  obtained
by the \texttt{Quantum ESPRESSO} distribution including \texttt{QE-PAW} datasets from the \texttt{PSlibrary}, 
for a large fraction of the periodic table,
including the alkaline-metals, the alkaline-earth and the $3d$ and $4d$
transition-metal elements (a total of 30 elements).

The CMST protocol reveals that for the majority of the systems tested \texttt{QE-PAW} and \texttt{Elk} LAPW show
excellent agreement in equilibrium lattice parameter within 0.4\%, in bulk modulus within 6\% and in energy differences 
within 1mRy, both in PBE and LDA, with the overall agreement slightly better for PBE. 
While performing the CMST, in the case of alkaline and alkaline-earth metals we have observed 
convergence and stability issues which point to a need for further studies that quantify the importance of 
the treatment of core electrons as well as the need of robust implementations. 

In the $\Delta$ factor tests, we have further extended the comparison between LAPW and PAW
for 68 elements in the periodic table, following the protocol recently proposed by Ref.\onlinecite{Lejaeghere2013}
{\it et al.}, based on a standardized study of the equation of state of the
elements in their experimental equilibrium phase. We have provided a second all-electron reference 
for this test set using the default values of \texttt{Elk} (unless a stability or converge issue has been encountered),
to represent a realistic scenario of the end-uses of all-electron methods,
and given that the agreement between \texttt{QE-PAW} and \texttt{Elk} has already been demonstrated by the CMST in the 
case of careful tuning. We have also extended the definition of the quality factor to $\Delta'$, 
a measure from which a proper distance satisfying the triangular inequality between codes/methods 
can be derived, allowing reference-independent comparisons.

For all flavors of the quality factor $\Delta$, and for the majority of the periodic table, 
we have found good agreement between the \texttt{QE-PAW} data, the \texttt{Elk} data, 
and the \texttt{WIEN2k} and \texttt{VASP} data from the literature.
This agreement highlights the reliability of the current state-of-the-art electronic structure 
codes/methods for a wide range of elements in the periodic table.
The differences observed for the remaining elements call attention to the 
urgent need of establishing reference data sets and standards for verification and validation
among computational approaches, both at the highest accuracy limit and for optimal, practical applications
by end-users. 

Finally, the extensive tests we provide 
make the \texttt{PSLibrary v0.3.1} a reliable default choice to be used within planewave pseudopotential implementations, 
that is openly available, updated and supported \cite{epfl_pseudo_repository}.

\begin{acknowledgments}
Two of us, BIA and GAA, are grateful to The Abdus Salam ICTP for hospitality and financial support 
as STEP Fellow and Regular Associate respectively, and acknowledge both ICTP and SISSA for computer time on their clusters.
\end{acknowledgments}


\begin{thebibliography}{99}

\bibitem{Onida2002}G. Onida, L. Reining and A. Rubio, Rev. Mod. Phys. {\bf 74}, 601 (2002).
\bibitem{Martonak2006}R. Martonak, D. Donadio, A. R. Oganov and M. Parrinello, Nature Materials {\bf 5} 623 (2006).
\bibitem{Curtarolo2013}S. Curtarolo, G. L. W. Hart, M. Buongiorno Nardelli, N. Mingo, S. Sanvito, and O. Levy, 
Nature Materials {\bf 12} 191 (2013). 
\bibitem{Jain2013}A. Jain, S.P. Ong, G. Hautier, W. Chen, W.D. Richards, S. Dacek, S. Cholia, 
D. Gunter, D. Skinner, G. Ceder and K.A. Persson, Applied Physics Letters Materials {\bf 1} 011002 (2013). 
\bibitem{Marzari2006}N. Marzari, MRS Bulletin {\bf 31} 681 (2006). 



\bibitem{NC} D. R. Hamann, M. Schl\"uter, and C. Chiang, Phys. Rev. Lett.
{\bf 43}, 1494 (1979).

\bibitem{US} D. Vanderbilt, Phys. Rev. B {\bf 41}, 7892 (1990).

\bibitem{Blochl1994} P.E. Bl\"ochl, Phys. Rev. B {\bf 50}, 17953 (1994).

\bibitem{Kresse1999} G. Kresse, D. Joubert, Phys. Rev. B {\bf 59}, 1758 (1999).

\bibitem{Curtiss1991}  L.A. Curtiss, K. Raghavachari, G.W. Trucks, and J.A. Pople, J. Chem. Phys. {\bf 94}, 7221 (1991).
\bibitem{Grossman2002} J. C. Grossman, J. Chem. Phys. {\bf 117}, 1434 (2002).
\bibitem{Nemec2010} N. Nemec, M.D. Towler, and R.J. Needs, J. Chem. Phys. {\bf 132}, 034111 (2010).

\bibitem{Paier2005}  J. Paier, R. Hirschl, M. Marsman, and G. Kresse, J. Chem. Phys. {\bf 122} 234102 (2005).

\bibitem{WIEN2kweb} P. Blaha,K. Schwarz, G. K. H. Madsen, D. Kvasnicka, J. Luitz: WIEN2k, 1999. 
“An augmented plane wave + local orbitals program for calculating crystal properties, Karlheinz Schwarz”. 
Austria: Techn. Universität Wien. ISBN:3-9501031 \texttt{http://www.wien2k.at}

\bibitem{GPAW}J. Enkovaara, C. Rostgaard, J. J. Mortensen, J. Chen, M. Dulak, L. Ferrighi, 
J. Gavnholt, C. Glinsvad, V. Haikola, H. A. Hansen, et al. J. Phys.: Condens. Matter {\bf 22}, 253202 (2010)
\texttt{https://wiki.fysik.dtu.dk/gpaw/}

\bibitem{VASPweb} G. Kresse and J. Hafner. 
Phys. Rev. B {\bf 49}, 14251 (1994). G. Kresse and J. Furthm\"uller. 
Comput. Mat. Sci. {\bf 6}, 15 (1996). G. Kresse and J. Furthm\"uller. 
Phys. Rev. B {\bf 54}, 11169 (1996). \texttt{https://www.vasp.at}

\bibitem{Lejaeghere2013} K. Lejaeghere, V. Van Speybroeck, G. Van Oost,
 and S. Cottenier, Critical Reviews in Solid State and Materials Sciences
{\bf 39}, 1 (2014);  arXiv:1204.2733.

\bibitem{GBRV} K.F. Garrity, J.W. Bennet, K.M. Rabe, and D. Vanderbilt,
Computational Materials Science {\bf 81}, 446-452 (2014); arXiv:1305.5973.

\bibitem{elkweb} \texttt{http://elk.sourceforge.net}

\bibitem{pslibrary} \texttt{http://www.qe-forge.org/gf/project/pslibrary}. Note that
among the tested elements the only differences between versions 0.3.0 and 0.3.1 occur 
for Ti, Cd, Ge, Te, Mn, Re, Po, Sb and Xe.  

\bibitem{espresso} P. Giannozzi, S. Baroni, N. Bonini, M. Calandra, R. Car, C. Cavazzoni, D. Ceresoli, 
G. L. Chiarotti, M. Cococcioni, I. Dabo, A. Dal Corso, S. Fabris, G. Fratesi, S. de Gironcoli, 
R. Gebauer, U. Gerstmann, C. Gougoussis, A. Kokalj, M. Lazzeri, L. Martin-Samos, N. Marzari, 
F. Mauri, R. Mazzarello, S. Paolini, A. Pasquarello, L. Paulatto, C. Sbraccia, S. Scandolo, 
G. Sclauzero, A. P. Seitsonen, A. Smogunov, P. Umari, R. M. Wentzcovitch, J. Phys.: Cond. Matter, {\bf 21}, 395502 (2009).
 \texttt{http://www.quantum-espresso.org/}
 %
\bibitem{Perdew1996} J.P. Perdew K. Burke and M. Ernzerhof, Phys. Rev. Lett. {\bf 77}, 3865 (1996).
\bibitem{Perdew1981} J.P. Perdew and A. Zunger, Phys. Rev. B {\bf 23}, 5048 (1981).
\bibitem{Madsen2001} Georg K.H. Madsen, Peter Blaha, Karlheinz Schwarz, Elisabeth Sjöstedt, 
and Lars Nordström Phys. Rev. B {\bf 64}, 195134 (2001).
\bibitem{meth} M.~Methfessel, and T.~Paxtox, Phys. Rev. B {\bf 40}, 3616 (1989). 
\bibitem{MV} N.~Marzari, D.~Vanderbilt, A.~De~Vita, and M.C.~Payne, Phys.\ Rev.\ Lett.\ {\bf 82}, 3296 (1999).
\bibitem{QEforge}   \texttt{http://www.qe-forge.org/} 
\bibitem{testrel} A. Dal Corso, Phys. Rev. B {\bf 86}, 085135 (2012).
\bibitem{Waleed} A. A. Adllan and A. Dal Corso, J. Phys.: Condens. Matter 
{\bf 23}, 425501 (2011).
\bibitem{PAWso} A. Dal Corso, Phys. Rev. B {\bf 82}, 075116 (2010).

%
\bibitem{Haas2009} P. Haas, F. Tran, and P. Blaha, Phys. Rev. B {\bf 79}, 085104 (2009).
\bibitem{Tran2007} F. Tran, R. Laskowski, P. Blaha, and K. Schwarz, Phys. Rev. B { \bf 75}, 115131 (2007).
\bibitem{Haas2009err} P. Haas, F. Tran, and P. Blaha, Phys. Rev. B {\bf 79}, 209902(E) (2009).
\bibitem{Ropo2008} M. Ropo, K. Kokko, and L. Vitos, Phys. Rev. B {\bf 77}, 195445 (2008).
\bibitem{Asato1999} M. Asato, A. Settels, T. Hoshino, T. Asada, S. Blügel, R. Zeller, and P. H. Dederichs, Phys. Rev. B {\bf 60}, 5202 (1999).
\bibitem{Mattsson2008} A. E. Mattsson, R. Armiento, J. Paier, G. Kresse, J. M. Wills, and T. R. Mattsson, J. Chem. Phys. {\bf 128}, 084714 (2008).
\bibitem{Sliwko1996} V.L. Sliwko, P. Mohn, K. Schwarz and P. Blaha, J. Phys.: Condens. Matter {\bf  8} 799.
\bibitem{Nobel1992}  J.A. Nobel and S. B. Trickey, P. Blaha and K. Schwarz, Phys. Rev. B {\bf 45}, 5012.
\bibitem{CMM-website} \texttt{http://molmod.ugent.be/DeltaCodesDFT}
\bibitem{recent} The Delta calculation package (version 2.0) distributed
on the CMM web-site\cite{CMM-website} and the data therein have
been used for the present calculations.
\bibitem{JTH} F. Jollet, M. Torrent, and N. Holzwarth. Comp. Phys. Commun. {\bf 185}, 1246 (2014).
{\sl "Generation of Projector Augmented-Wave atomic data:
a 71 elements validated table in the XML format"}, arXiv:1309.7274v2 




\bibitem{dalcorso100} A. Dal Corso, to be published.
\bibitem{epfl_pseudo_repository} Pseudopotential repository at http://theossrv1.epfl.ch/Main/Pseudopotentials







 




\end{thebibliography}
\end{document}